\documentclass[12pt,preprint]{aastex}

\shorttitle{Failed Supernovae}
\shortauthors{Fryer et al.}

\def \nuc#1#2{\relax\ifmmode{}^{#1}{\protect\text{#2}}\else${}^{#1}$#2\fi}

\begin{document}

\title{Spectra and Light Curves of Failed Supernovae}

\author{Chris L. Fryer\altaffilmark{1,2}, Peter
  J. Brown\altaffilmark{3}, Filomena Bufano\altaffilmark{4}, Jon A.
  Dahl\altaffilmark{1}, Christopher J. Fontes\altaffilmark{1}, Lucille
  H. Frey\altaffilmark{1}, Stephen T. Holland\altaffilmark{5}, Aimee
  L. Hungerford\altaffilmark{1}, Stefan Immler\altaffilmark{5}, Paolo
  Mazzali\altaffilmark{6,7,8}, Peter A. Milne\altaffilmark{9}, Evan
  Scannapieco\altaffilmark{10}, Nevin Weinberg\altaffilmark{11}, Patrick
  A. Young\altaffilmark{10}}

\altaffiltext{1}{Los Alamos National Laboratory, Los Alamos, NM
87545}

\altaffiltext{2}{Physics Dept., University of Arizona, Tucson AZ
85721}

\altaffiltext{3}{Pennsylvania State University, Dept. of Astronomy 
\& Astrophsycis, University Park, PA 16802}

\altaffiltext{4}{Dipartimento di Astronomia, Univ. Padova,
  INAF-Osservatoroio Astronomico di Padova}

\altaffiltext{5}{Astrophysics Science Division, NASA Goddard Space 
Flight Center, Greenbelt, MD 20771}

\altaffiltext{6}{Max-Planck-Institut f\"ur Astrophysik,
Karl-Schwarzschild-Str. 1, 85748 Garching, Germany}

\altaffiltext{7}{Padova: INAF-Osservatorio Astronomico di Padova, 
Padova, Italy}

\altaffiltext{8}{Scuola Normale Superiore, Piazza dei Cavalieri, 7, 
56126 Pisa, Italy}

\altaffiltext{9}{Steward Observatory, 933 North Cherry Avenue, RM
N204, Tucson, AZ 85721}

\altaffiltext{10}{SESE, Arizona State University, Tempe, AZ 85287}

\altaffiltext{11}{Astronomy Department and Theoretical Astrophysics
  Center, University of California, Berkeley, 601 Campbell Hall,
  Berkeley, CA 94720, USA}

\email{fryer@lanl.gov, grbpeter@yahoo.com,
  filomena.bufano@oapd.inaf.it, dahl@lanl.gov, cjf@lanl.gov,
  aimee@lanl.gov, sholland@milkyway.gsfc.nasa.gov,
  stefan.m.immler@nasa.gov,
  mazzali@MPA-Garching.MPG.DE,pmilne511@cox.net,
  evan.scannapieco@asu.edu, nweinberg@astro.berkeley.edu,
  patrick.young.1@asu.edu}

\begin{abstract}

Astronomers have proposed a number of mechanisms to produce supernova
explosions.  Although many of these mechanisms are now not considered
primary engines behind supernovae, they do produce transients that
will be observed by upcoming ground-based surveys and NASA satellites.
Here we present the first radiation-hydrodynamics calculations of the
spectra and light curves from three of these ``failed'' supernovae:
supernovae with considerable fallback, accretion induced collapse of
white dwarfs, and energetic helium flashes (also known as type .Ia
supernovae).

\end{abstract}


\section{Introduction}

Supernovae (SNe) and gamma-ray bursts (GRBs) are among the brightest
transients in the universe.  As such, they have been well-studied,
both observationally and theoretically.  Although many theoretical
models have been proposed to explain the engines behind these
explosions, astronomers have focused on a few, best-fitting
``standard'' models.  The rest of the models were, for the most part,
discarded because either the engine, when studied in more detail,
could not explain the observed SN characteristics (e.g. the explosion
was weaker than that needed to explain most supernovae or gamma-ray
bursts) and/or the rate of explosions was below the observed SN or GRB
rate.

These ``failed''\footnote{Explosions that don't produce the standard 
models for supernovae.} supernovae have been neglected: very few studies have
focused on their explosions and virtually no studies have calculated
the emission from the explosions.  With their typically dimmer
outbursts and often lower rates, these objects were unlikely to have a
large presence in past supernova surveys.  But current (Palomar
Transient Factory, PanSTARRS) and upcoming (SkyMapper, VLT Survey
Telescope, One Degree Imager, Large Synoptic Survey Telescope)
transient surveys are likely to actually observe these neglected
transients.  In this paper, we present some of the first spectra and
light curves from radiation-hydrodynamics models of a few of these
transients to both help guide searches and use the observations of the
transients to constrain our understanding of the explosions.

The emission from explosions is powered by two primary sources: the
decay of radioactive elements produced in the explosion and shock
heating as the ejecta blows through the medium surrounding it.  These
two energy sources play varying roles in supernovae and GRBs.  We
believe that shock heating (albeit through shock acceleration
mechanisms) powers the emission in GRBs (e.g. \cite{MR92}).
\cite{FYH06} found that even the GRB-associated supernova could well
be dominated by shock heating.  For supernovae, the dominant energy
source depends on the class, or type, of supernovae: the decay of
radioactive $^{56}$Ni and its daughter products dominates the type Ia
emission but for many type Ib/c and II supernovae, shock heating can
dominate the emission at peak (e.g. Frey et al. 2009, in preparation).
The radiation-hydrodynamics calculations in this paper allow us to
include both power sources and determine the crucial conditions behind
the observed emission of these explosions.

Many transients also are sources of gravitational wave (GW) and
neutrino emission.  They exhibit different features form normal
supernovae and these differences can be used to help us understand
both explosion mechanisms better.  Neutrino and GW observations of
failed supernovae provide complementary (and, in some cases, stronger)
probes of nuclear physics and general relativity.

In this paper, we study 3 ``failed'' supernova models: accretion
induced collapse (AIC) of a white dwarf (Chapter 3), dim supernovae
produced by fallback (Chapter 4), and type .Ia supernovae (Chapter 5).  We
review the engine and its environment, estimate the occurrence rate,
show spectral and light-curve results from radiation-hydrodynamics
calculations (using the RAGE supernova emission code---see Chapter 2) of
these explosions, and discuss the neutrino and GW emission for each of
these explosions.

\section{Code Description}
\label{sec:code}

To include shock heating in our light-curve calculations, we must
couple our radiation transport calculation with a hydrodynamics
package.  For our radiation-hydrodynamics calculations, we use the
multidimensional radiation-hydrodynamics code RAGE (Radiation Adaptive
Grid Eulerian), which was designed to model a variety of multimaterial
flows \citep{Bal96}.  The conservative equations for mass, momentum,
and total energy are solved through a second-order, direct-Eulerian
Godunov method on a finite-volume mesh \citep{Git08}.  It includes a
flux-limited diffusion scheme to model the transport of thermal
photons using the Levermore-Pomraning flux limiter \citep{LP81}.  RAGE
has been extensively tested on a range of verification problems
\citep{Hol99,Hue05} and applied to (and tested on) a range of
astrophysics problems~\citep{Her06,Cok06,Fry07,FHY07}, including the
strong velocity gradients that exist in supernova
explosions~\citep{LR06}.

The RAGE code can be used in 1, 2, and 3 dimensions with spherical,
cylindrical and planar geometries in 1-dimension, cylindrical and
planar geometries in 2-dimensions, and planar geometries in
3-dimensions.  For this paper, we limit our analysis to 1-dimensional,
spherical calculations.  RAGE uses an adaptive mesh refinement
technique, allowing us to focus the resolution on the shock and follow
the shock as it progresses through the star.  Even so, we were forced
to regrid in the calculations to ensure that the shock was resolved (typically 
with coarse-grid cell sizes set to a few percent of the shock position and 
fine grid cell sizes set to a fraction of a percent) at
early times but still allow us to model the shock progression out to
40--100\,d (the shock moves from $10^9$\,cm out to $10^{16}$\,cm in the
course of a simulation).

For most of our calculations, the energy released from the 
decay $^{56}$Ni and $^{56}$Co is deposited directly at the location 
of the $^{56}$Ni using the following formula:
\begin{equation}
dE/dt = E_{\rm Ni}/\tau_{\rm Ni} e^{-t/\tau_{\rm Ni}} +
E_{\rm Co}/(\tau_{\rm Co} - \tau_{\rm Ni}) 
[ e^{-t/\tau_{\rm Co}} - e^{-t/\tau_{\rm Ni}} ]
\end{equation}
where $E_{\rm Ni}=1.7$\,MeV and $E_{\rm Co}=2.9$\,MeV are the mean
energies released per atom for the decay of $^{56}$Ni and $^{56}$Co,
respectively, and $\tau_{\rm Ni}=7.6 \times 10^5$\,s, $\tau_{\rm
Co}=9.6 \times 10^6$\,s.  Especially at late times, this energy is not
deposited into the matter surrounding it, but rather escapes the star.

In order to test the accuracy of the assumption of in-situ energy deposition,
we have run a single simulation including the transport of the
gamma-rays emitted during the decay of $^{56}$Ni and its daughter
product $^{56}$Co.  These results are compared with our local deposition
models.  To solve the transport equation in this calculation, we use
the discrete ordinates ``$S_N$'' method (Wick 1943; Chandrasekhar
1950, Carlson 1955).  In the spherical, 1-dimensional calculations used
here, we discretize the angular variables as:
\begin{eqnarray}
&& \frac{1}{c} \frac{\partial I_n(r,\nu)}{\partial t} + \frac{\mu_n}{r} \frac{\partial r^2 I_n(r,\nu)}{\partial
r} + \frac{2}{r w_n} \left[\alpha_{n+1/2} I_{n+1/2}(r,\nu) - \alpha_{n-1/2} I_{n-1/2}(r,\nu)\right] + \\ 
\nonumber && \sigma_{\rm tot}(r,\nu) I_n(r,\nu)  =  
 \int_0^\infty d\nu \sum_{l=0}^L (2l+1) \sigma_{\rm scat,l}(r,\nu) P_l(\mu_n) \sum_m P_l(\mu_m) I_m(r,\nu) w_m + Q_n(r,\nu)
\end{eqnarray}
where $c$ is the speed of light, $I_n(r,\nu)$ is the angular intensity
as a function of space coordinate $r$ and photon energy $\nu$,
$\mu_n = \cos \theta_n$ is the discretized $\mu$ and is taken from the
abscissas of the standard 1-dimensional Gauss Legendre quadrature with
$w_n$ the weights of this quadrature, $\alpha_{n+1/2} = \alpha_{n-1/2}
- \mu_n w_n$ is the angular differencing coefficient (with
$\alpha_{1/2}=0$), $\sigma_{\rm tot}(r,\nu)$ is the macroscopic total
cross-section, $\sigma_{\rm scat,l}(r,\nu)$ is the $l^{\rm th}$
Legendre moment of the differential scattering cross-section,
$P_l(\mu_n)$ is the Legendre polynomial of $l^{\rm th}$ order and
$Q_n(r,\nu)$ is our discretized source arising from the radioactive
decay.  The energy-dependent variable is discretized using standard
multi-group theory (we use 12 groups).  Spatial and angular cell edges
are related to their respective cell centers by the standard diamond
difference approach and time integration is done using
Crank-Nicholson.  The transport operator is inverted using a
space-angle sweep, one energy group at a time.  The multi-group
cross-section data comes from the Los Alamos MENDF6 library~\citep{Lit96}.

\begin{figure}
\plottwo{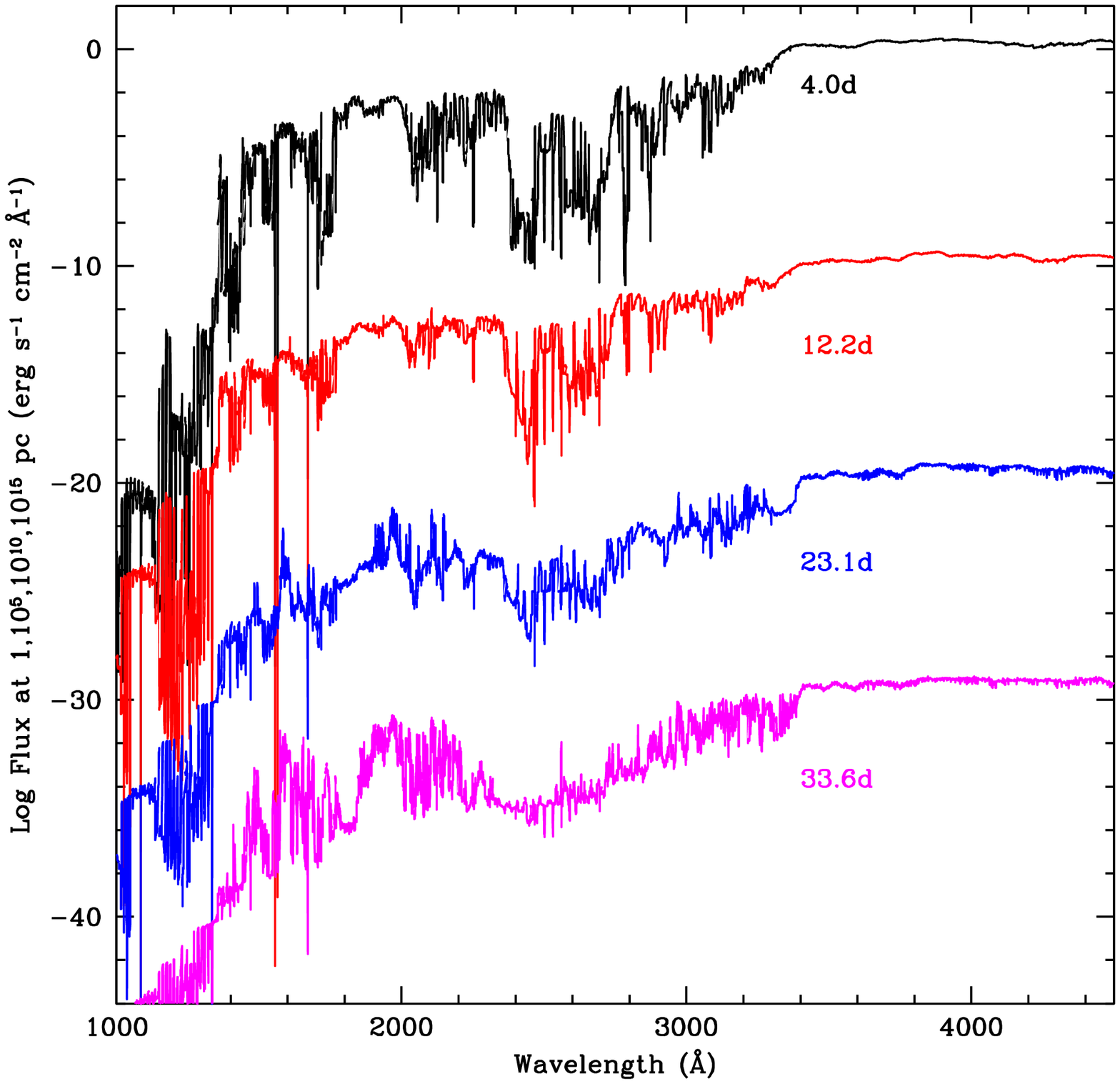}{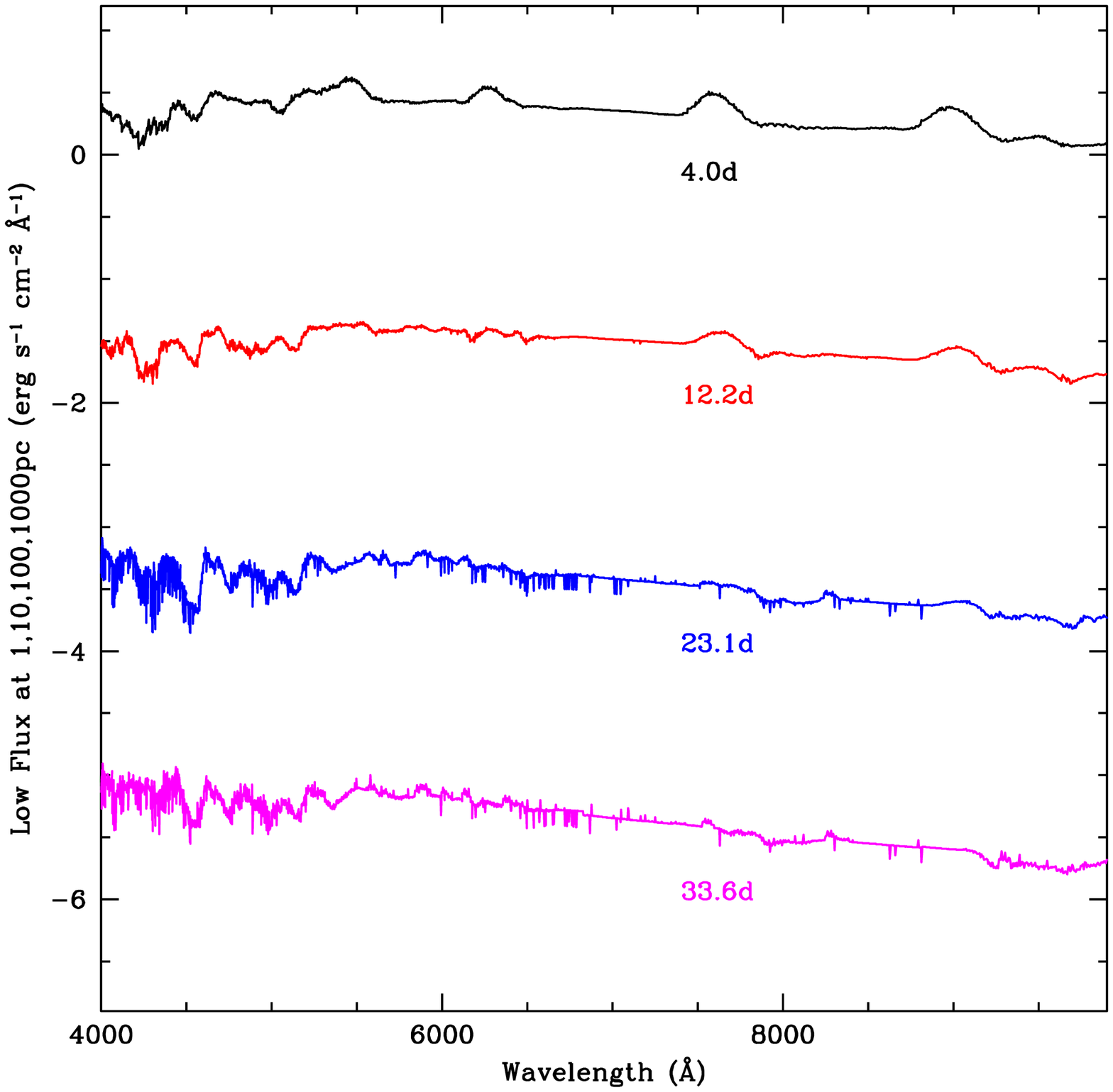}
\plottwo{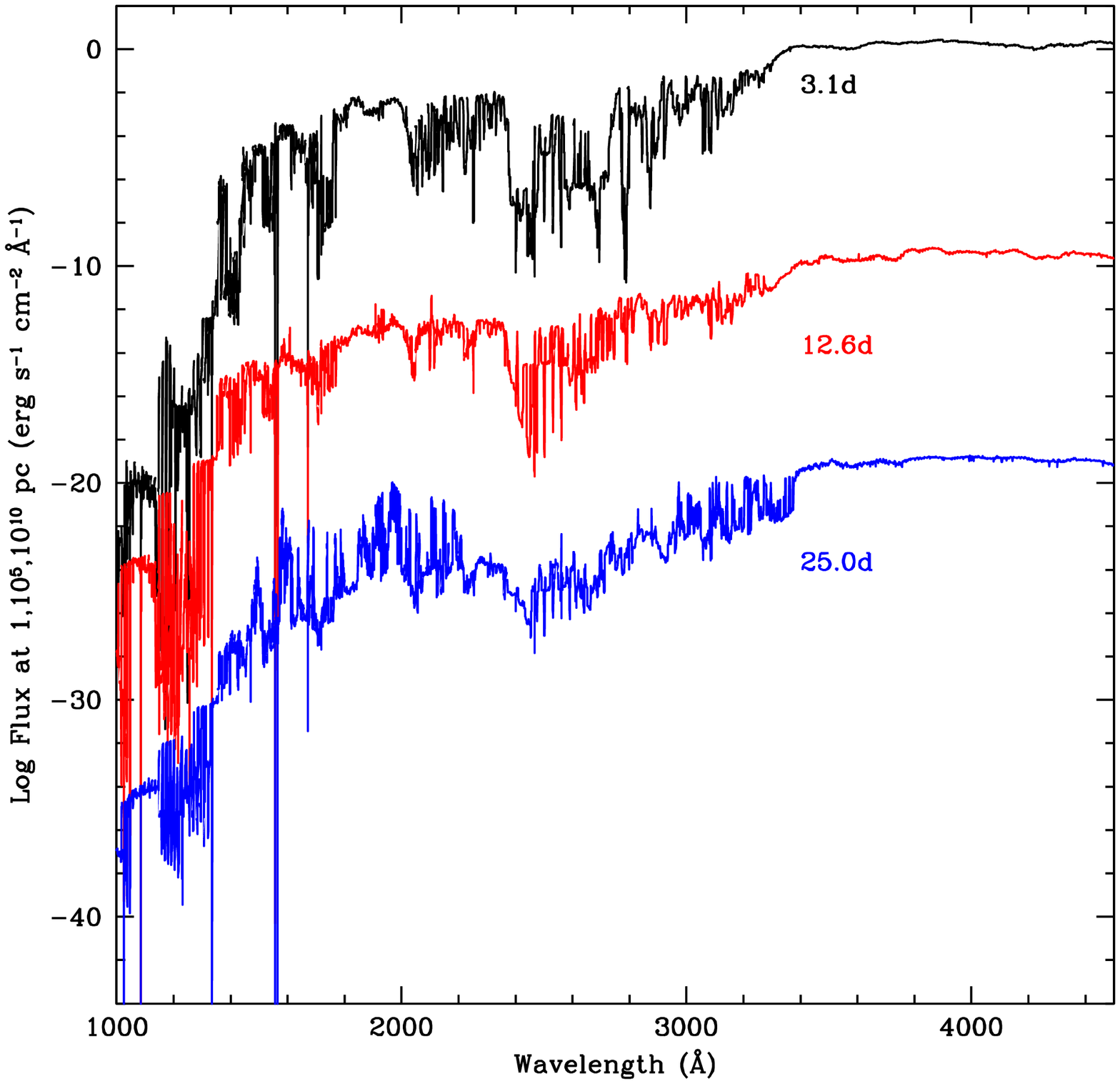}{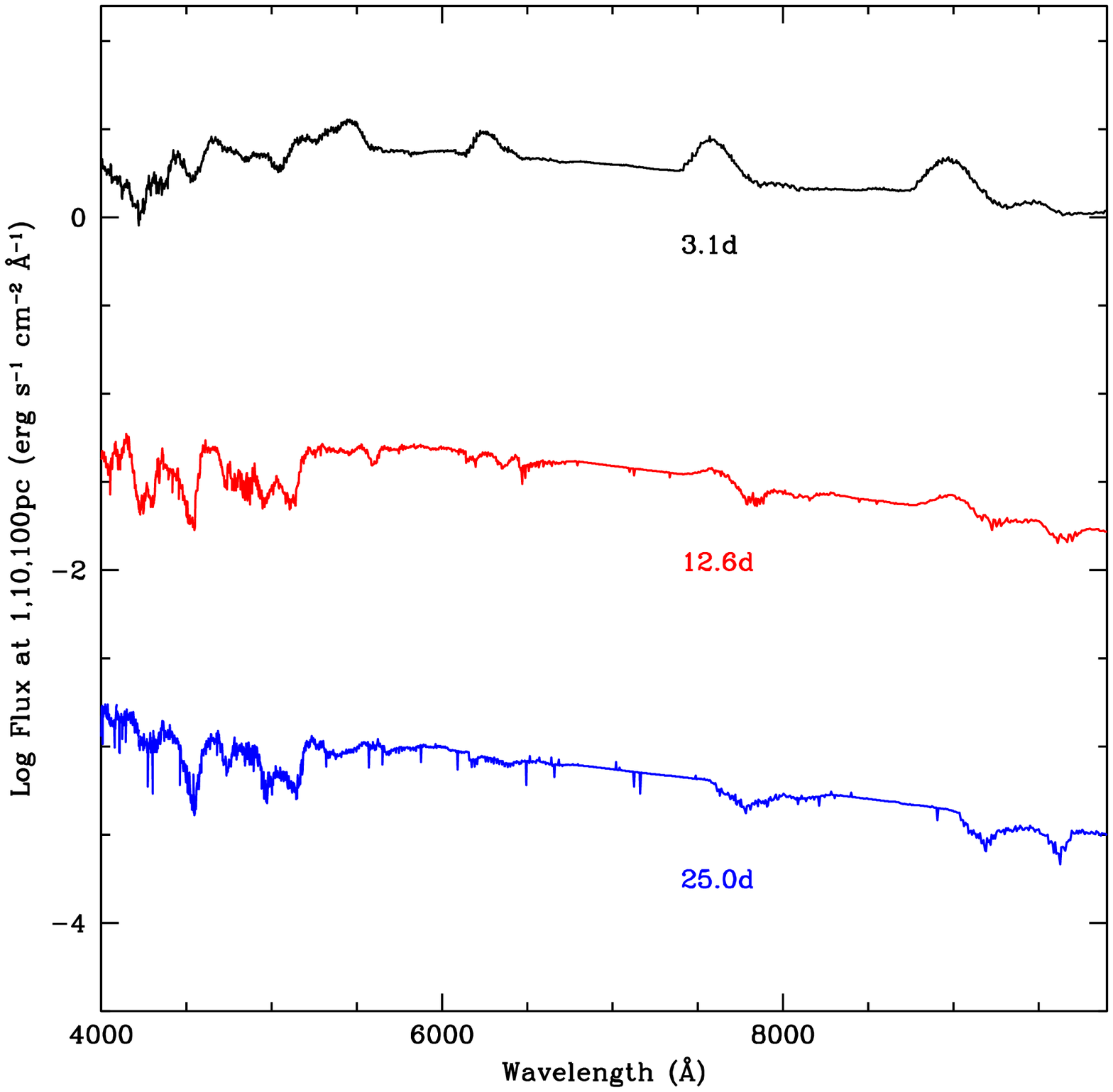}
\caption{Flux versus wavelength for a 16\,M$_\odot$ binary progenitor
  (collapsing as a Wolf-Rayet star of less than 5\,M$_\odot$) at 4
  different times after the launch of the explosion.  The high energy
  spectra are separated each by 10 orders of magnitude to show the
  full structure of each spectrum.  We have effectively put the
  different models and different distances: 1\,pc, 10$^5$\,pc,
  10$^{10}$\,pc and 10$^{15}$\,pc.  The corresponding fluxes at low
  energy, which have much less dramatic structures, are separated by 2
  orders of magnitude.  The top two panels show the spectra for
  in-situ gamma-ray deposition.  The bottom two panels show the
  resulting spectra using our gamma-ray transport algorithm.  At these
  times, the differences are minimal.  Indeed, our tests show that
  even for a type Ia supernova, gamma-ray transport is not critically
  important until well after 60\,d.}
\label{fig:16Hspec}
\end{figure}

For the comparison of our in-situ gamma-ray deposition to that of
transport, we model a Wolf-Rayet star.  The spectra for our in-situ
gamma-ray deposition and transported gamma-ray calculations are shown
in Figure~\ref{fig:16Hspec}.  At these early times, the two
calculations are identical.  The mean free path of gamma-rays remains
small for this model, and most of the models studied in this paper, for
roughly 60\,d (longer for some), so the fact that in-situ deposition
is a good approximation is not surprising.  The one exception is our 
low density .Ia model.  In this .Ia model, the gamma-rays emitted 
by the decay $^{56}$Ni are not trapped after $\sim$15\,d.  We discuss 
these results further in section 5.

For opacities, our radiation hydrodynamics calculations consider a
single group using the Rosseland mean opacity for the diffusion
coefficient and the Planck mean opacity for the emission/absorption
terms in the transport equation\footnote{We have run one calculation
  in our fallback runs using 5 groups (see section 4).  Although the
  basic fluxes remain the same, the spectral line strengths will
  vary.}.  These gray opacities are obtained from the LANL OPLIB
database (Magee et al. 1995;
\url{http://www.t4.lanl.gov/cgi-bin/opacity/tops.pl}) and have been
extensively used in astrophysics modeling, including many problems in
supernovae (e.g. Fryer et al. 1999b; Deng et al.  2005; Mazzali et al.
2006). This opacity database is continually updated, and we use the
most recently produced opacity data in all of our calculations. The
opacities made available in this database are computed under the
assumption that the atomic populations are in local thermodynamic
equilibrium at the material temperature. Thus, the opacity can be
determined assuming a single temperature in each cell.

\begin{figure}
\plottwo{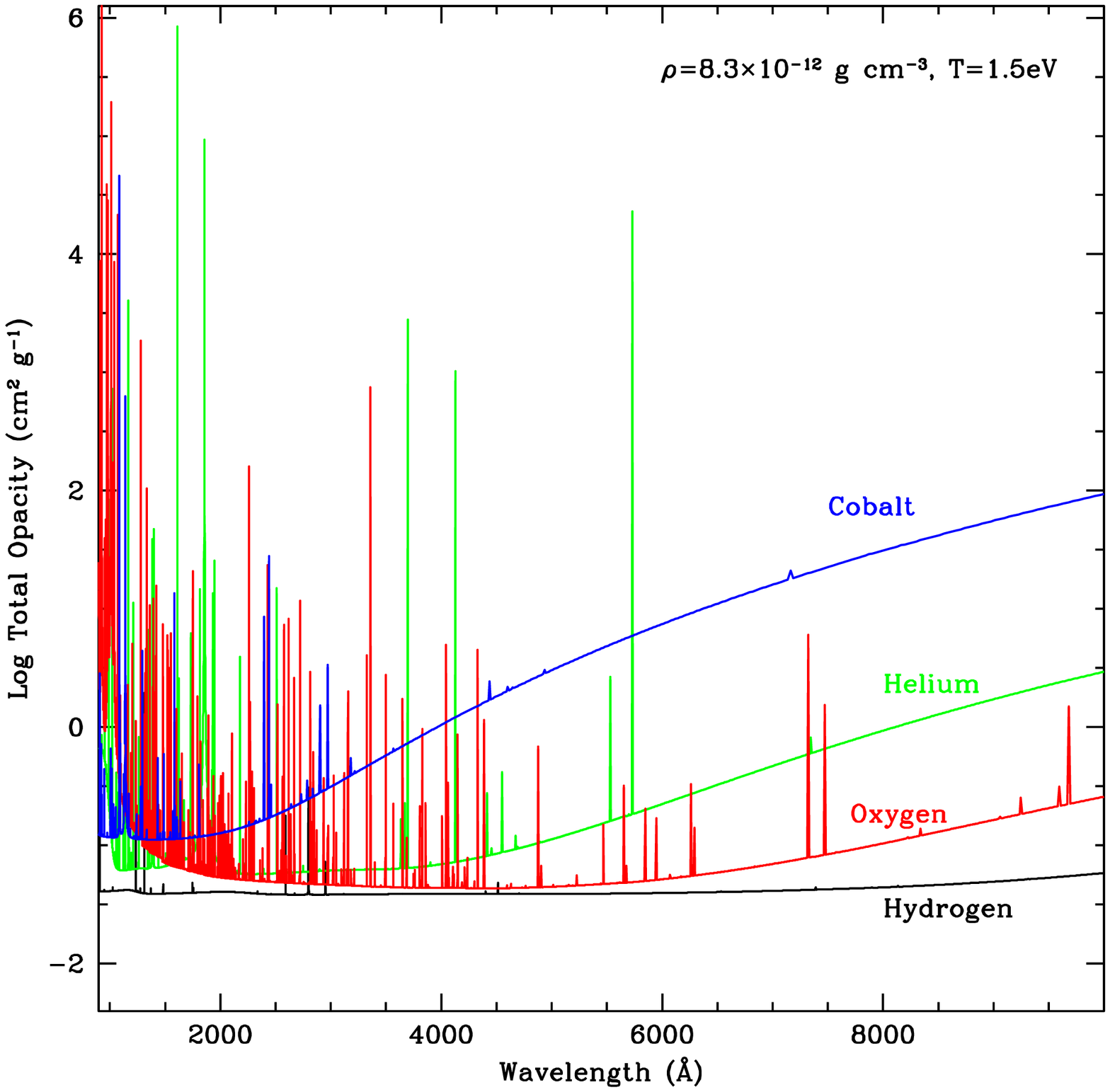}{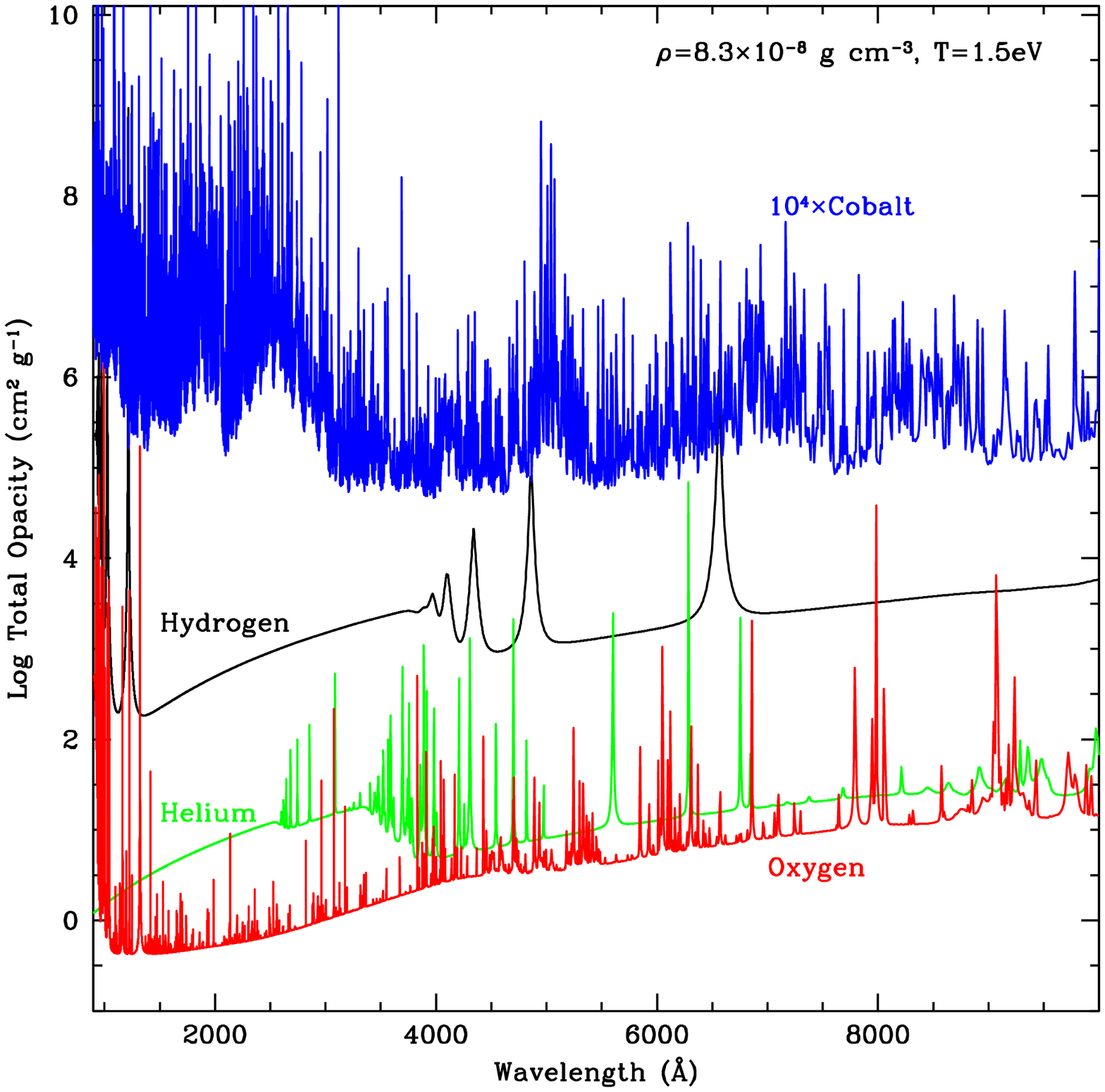}
\epsscale{0.5}
\plotone{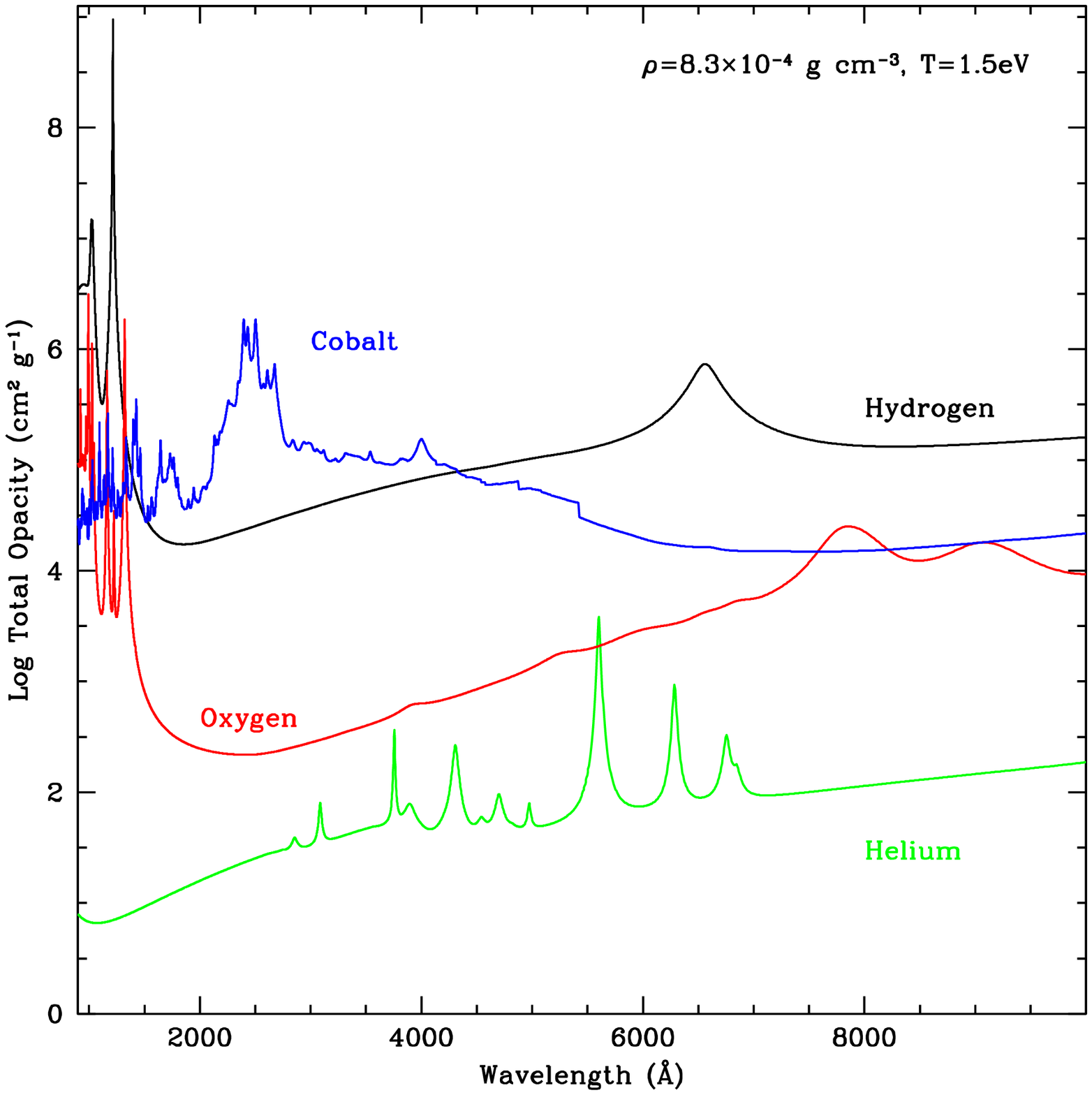}
\epsscale{1.0}
\caption{Opacities for hydrogen, helium, oxygen and cobalt versus
wavelength for 3 different density/temperature pairs relevant to our
supernova explosion.  The high density plot
($8.3\times10^{-4}$\,g\,cm$^{-3}$)
is a typical density near the time of shock break-out.
Under these conditions, the hydrogen opacity is quite large (3 orders
of magnitude higher than the equivalent helium opacity) and many lines 
are blended for each material.  At lower densities,
$8.3\times10^{-8}$\,g\,cm$^{-3}$,
the lines dominate the opacities.  The Doppler shifts in 
the exploding star are easily sufficient to cause these lines to 
blend.  At still lower densities, the opacities are 
quite low and the iron peak material has considerably fewer lines.}
\label{fig:opac}
\end{figure}

\begin{figure}
\plottwo{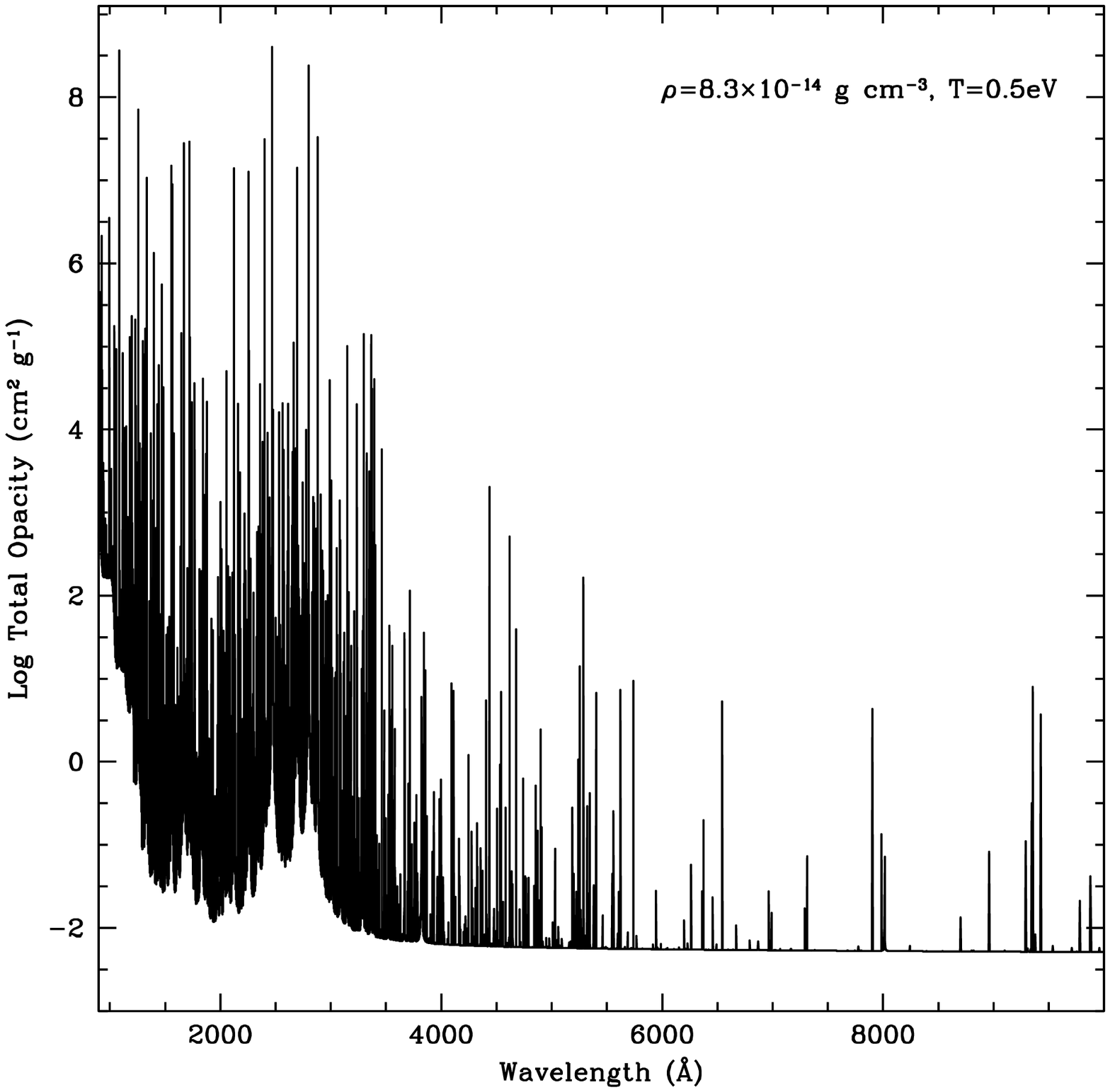}{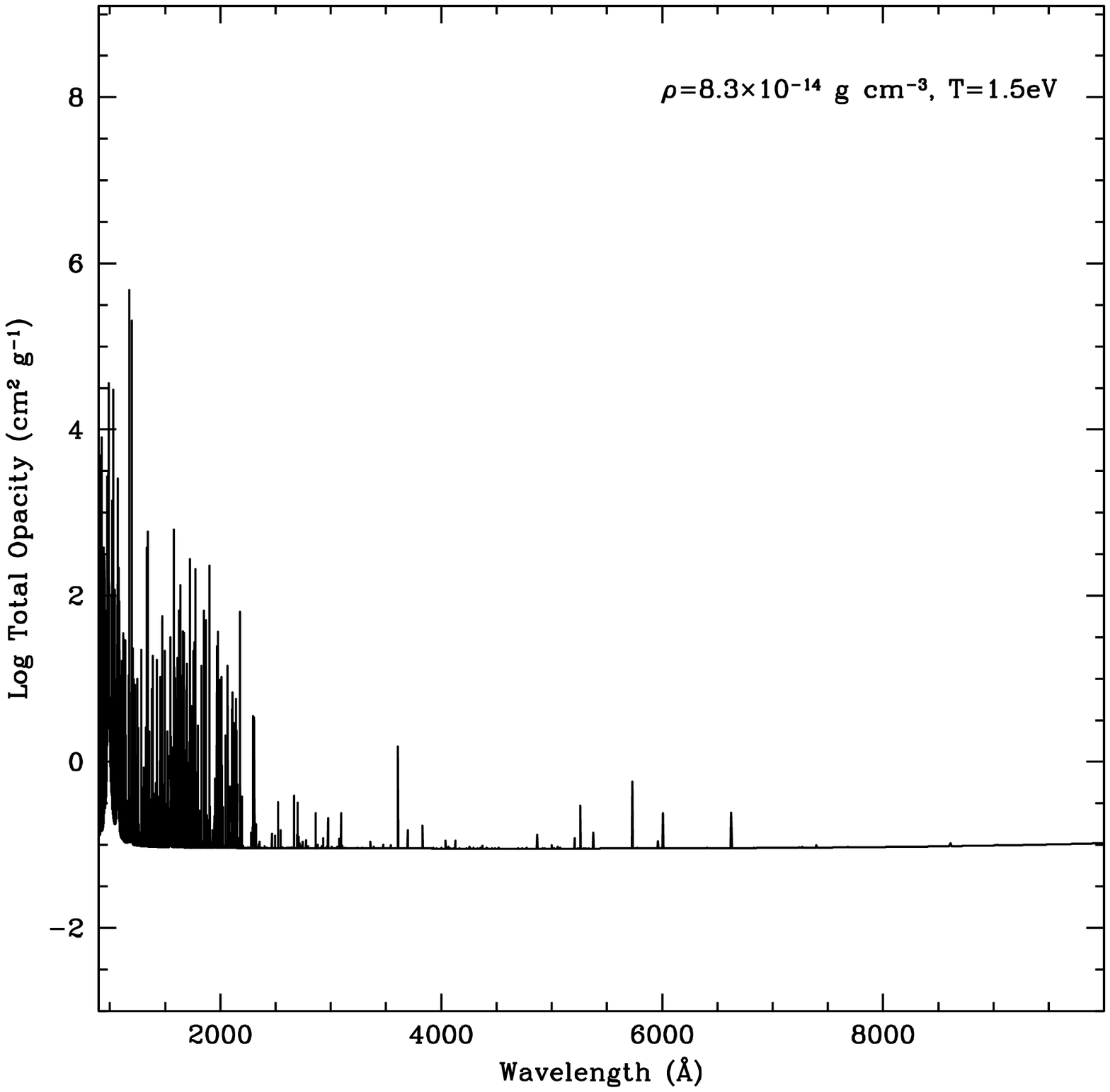}
\plottwo{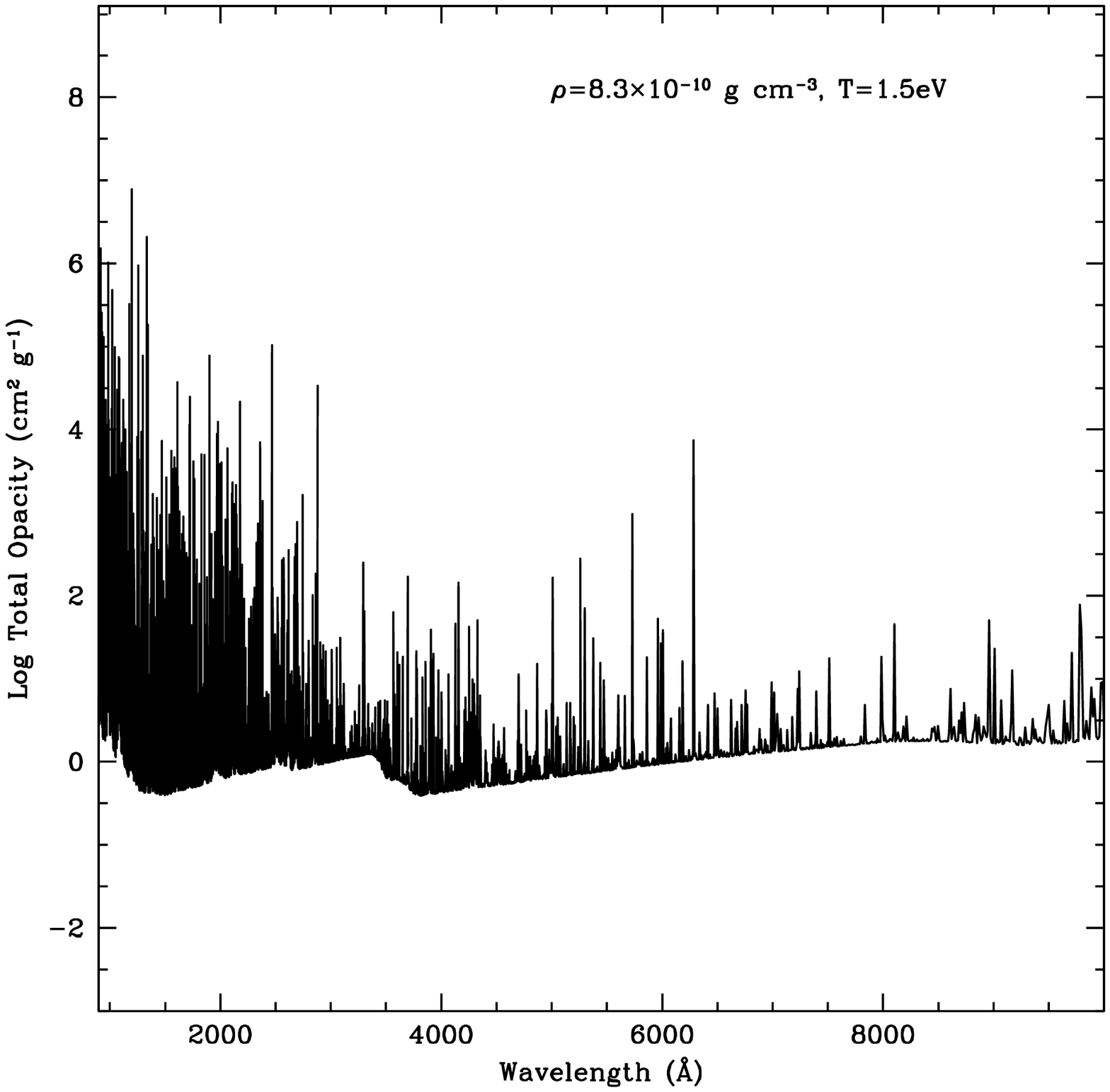}{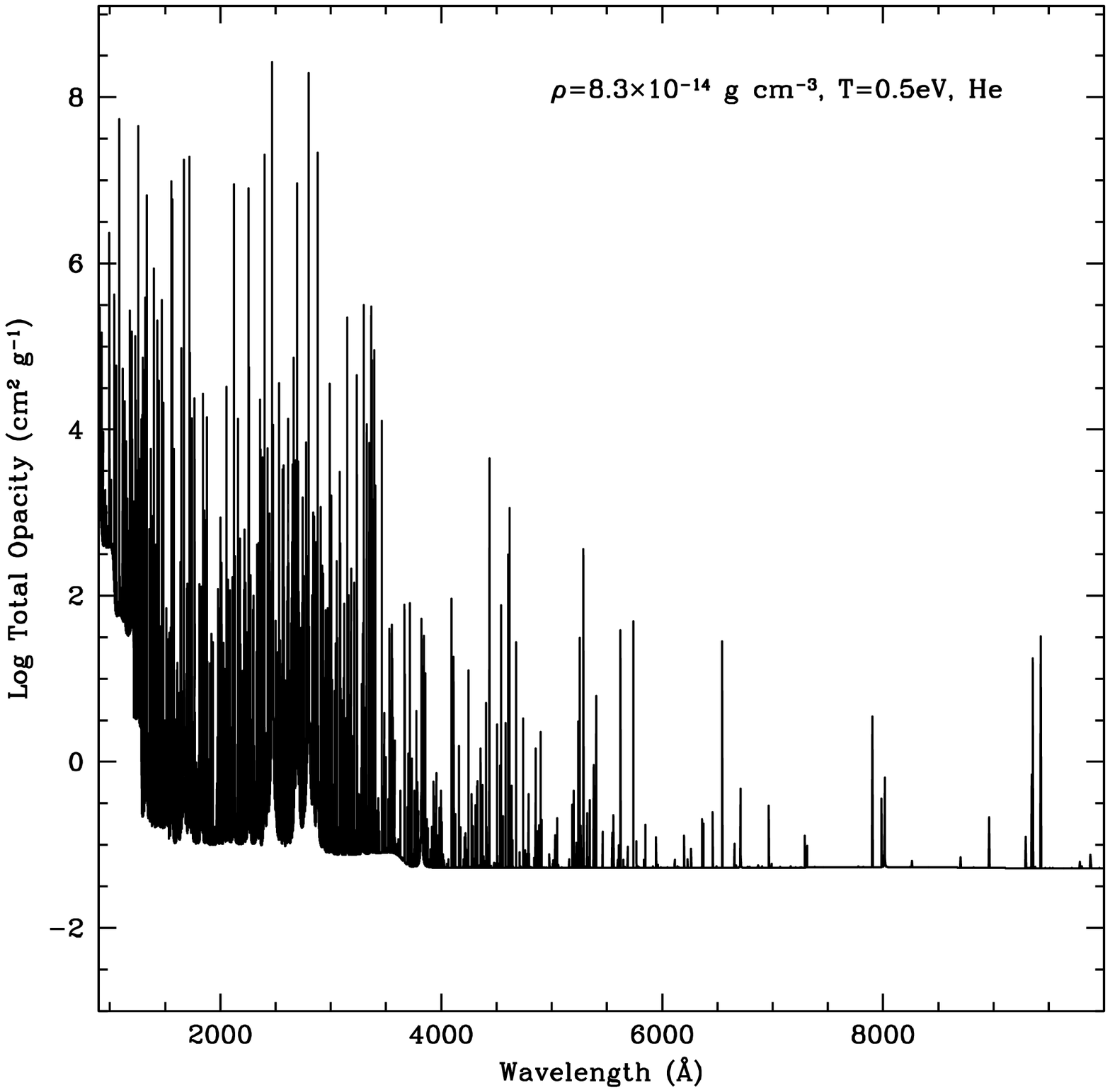}
\caption{Opacities for our wind mixture used in our fallback
  calculations at 3 different density/temperature pairs relevant to
  our supernova explosion.  Note that the placement of the lines can
  move dramatically based on the temperature profile.  We have also
  included one plot showing the opacity for a mixture where the
  hydrogen has been replaced by helium.}
\label{fig:opac2}
\end{figure}

As an illustration of the opacities used in our calculations, we have
plotted the opacity values from the LANL OPLIB database for a variety
of pure elements at 3 different density/temperature pairs
(Fig.~\ref{fig:opac}).  At low densities, we note that hydrogen in
local thermodynamic equilibrium will be completely ionized, even at
temperatures as low as 1eV, because the effect of three-body
recombination is suppressed relative to that of photoionization.
Thus, bound-bound features associated with the hydrogen atom, such as
the H$\alpha$ line, are not expected to be present under these
conditions.  The pure elemental opacities are subsequently combined in
the appropriate ratios for each cell that is considered in the
calculation.  Figure~\ref{fig:opac2} shows 3 different
density/temperature pairings for the composition of the surrounding
wind medium used in our fallback model.

With our radiation-hydrodynamics calculations, we calculate the
temperature structure of the matter in the exploding star as a
function of time.  Unlike post-process calculations based on purely
hydrodynamic models, we can use this matter temperature profile in
a post-process approach to
determine the full time-dependent spectra from this supernova
explosion.  To calculate the spectrum, we first assume that each 
radial zone emits radiation isotropically based on its temperature 
and absorption coefficient:
\begin{equation}
L_\nu = m_{\rm zone} \frac{2 h \sigma_{\rm abs} \nu^3}{c^2} \frac{1}{e^{\nu/T_{\rm mat}}
- 1} d\nu \frac{(1-v/c)^2}{\sqrt{1-(v/c)^2}}
\end{equation}
where $m_{\rm zone} $ is the mass of the zone, $h=6.626 \times
10^{-27} {\rm erg s}$ is Planck's constant, $c$ is the speed of light,
$v$ is the velocity of that zone, $(1-v/c)^2/\sqrt{1-(v/c)^2}$ is the
time dilation effect on the luminosity, $\sigma_{\rm abs}$ is the
absorption cross-section (which depends on composition, temperature and
density), $\nu$ is the frequency ($d\nu$ is the size of the frequency
bin), and $T_{\rm mat}$ is the matter temperature (note that in the
exponential, $\nu$ and $T_{\rm mat}$ must have the same units---e.g. $h
\nu/k_{\rm Boltzmann} T_{\rm mat}$).

\begin{figure}
\epsscale{0.7}
\plotone{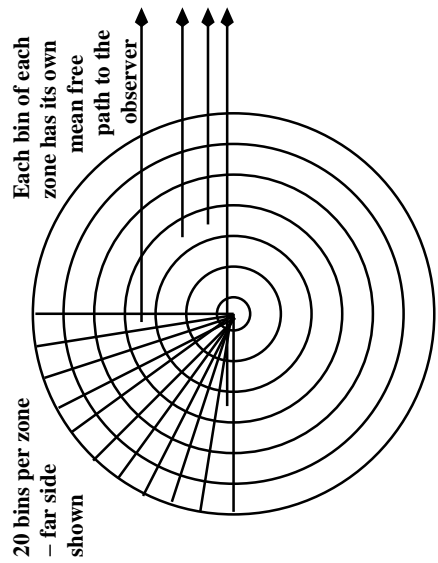}
\epsscale{1.0}
\caption{Our post-process code calculates the photons assuming one
viewing angle, calculating the unabsorbed photons.  A given zone will
contribute material moving toward and away from that viewing angle.
Each zone is broken into a number of angular bins to calculate
this red and blue shift.  The post-process approach to transport assumes that
the transport time of unabsorbed photons is short compared to the
hydrodynamic timescale.}
\label{fig:postprocess}
\end{figure}

This equation gives us the emission in each zone, but what we really
want is the emission directed toward an observer in a single
direction.  In order to calculate both accurate mean free paths through the
spherically symmetric zones (to get limb effects) and the
correct Doppler shifts, we have discretized each zone into angular
bins (Fig.~\ref{fig:postprocess}).  For our calculations, we use 40
angle bins.  The observed spectrum is then:
\begin{equation}
L^{\rm tot}_\nu = \sum_{\rm zone} \sum_{\rm angle} L_\nu^{{\rm angle,zone}} e^{-\tau^{{\rm angle,zone}}}
\end{equation}
where $\tau^{\rm angle,zone}$ includes both Doppler effects (everything is 
calculated in the rest frame of the observer) and geometric or limb 
effects.  $L_\nu^{\rm angle,zone}$ is now the emission based on the mass 
in our angular bin ($m_{\rm zone,l}$) pointed in the observers direction:
\begin{equation}
L_\nu = \frac{m_{\rm angle,zone}}{n_{\rm angular \, bins}} \frac{2 h \sigma_{\rm abs}
\nu^3}{c^2} \frac{1}{(e^{\nu/T_{\rm mat}} - 1)} d\nu \frac{(1-v/c)^2}{\sqrt{1-(v/c)^2}}
\end{equation}
where $n_{\rm angular \, bins}=40$ in our case.  As long as our
assumption holds concerning the accuracy of the matter temperature
obtained from our radiation-hydrodynamics calculation, this
semi-analytic post-process gives us an accurate calculation of
the emission.  We then can calculate the emission over our entire
energy grid consisting of 14,900 groups from roughly $10^{-3}$\,eV to
$10^{4}$\,eV (the grid depends upon the temperature and density of 
the matter).  For typical supernova temperatures and densities, we 
generally have $\sim 13,000$ groups lying between 1000 and 10,000\,\AA.

To obtain optical and UV light curves over the wavelength range
1600-6000\AA, , we need to integrate our spectrum over a band filter.
In our case, we use the SWIFT band filters for U, B, and V (Gehrels et al 2004; 
Roming et al. 2005; Poole et al. 2008): swuuu\_20041120v104.arf,
swubb\_20041120v104.arf, and swuvv\_20041120v104.arf to be specific.
We also include the data for the swuw1\_20041120v104.arf,
swuw2\_20041120v104.arf, and swum2\_20041120v104.arf UV bands.

\section{Accretion Induced Collapse}
\label{sec:aic}

When accretion onto a white dwarf pushes its mass above the
Chandrasekhar limit, the star begins to compress.  This compression
can lead to one of two fates.  In one scenario, nuclear burning
releases enough energy to completely unbind the star in a
thermonuclear explosion, producing the well-known type Ia supernova
used to probe the early universe.  In the other, the white dwarf
collapses down to a neutron star (accretion induced collapse
or AIC).  The gravitational potential energy released in this collapse
also produces an explosion.  It is this latter, lesser-known,
accretion induced collapse that we study here.  An accretion induced
collapse can only form if nuclear burning during the collapse does not
inject enough energy to unbind the star.  If the core of the white
dwarf is cool enough such that nuclear burning does not occur (or is
weak) until after the core has imploded (and lost energy through
neutrino emission), nuclear burning will be unable to unbind the star.

Nomoto \& Kondo (1991) summarized the fate of an accreting white dwarf
based on its composition (carbon-oxygen versus oxygen-magnesium-neon
white dwarf), initial mass, and accretion rate.  They argued that
white dwarfs with initial masses above 1.2M$_\odot$ are likely to form
AICs.  Unless the accretion rate is quite low
($<10^{-6}$M\,$_\odot$\,y$^{-1}$), the mass of these white dwarfs will
exceed the Chandrasekhar mass well before accretion energy can heat
the core.  The fact that many of these massive white dwarfs are OMgNe
white dwarfs whose cores are cooled by Urca processes (the emission of
a neutrino and anti-neutrino pair within a nucleus) does not help.
They also argued that AICs are formed when the accretion rate is high,
again causing the white dwarf mass to exceed the Chandrasekhar limit
before the core is heated by this accretion.  With these two
constraints, we can study the rates of AICs.

Note that an accretion induced collapse has many properties similar to
that of electron capture supernovae.  An electron capture supernova is
produced in an AGB star with a mass placing its evolution at the
boundary between mass ejection (forming a white dwarf) and further
core nuclear burning producing an iron core and, ultimately, a iron
core-collapse supernovae (see \cite{Whe98},\cite{Wan03}, and
\cite{Poe08}).  An electron capture supernova is produced in the
collapse of the OMgNe core at the center of this AGB star.  The
details of the explosion for an electron capture supernovae are very
similar to those of an AIC.  The primary difference between these
objects is the surrounding environment and, as we shall see, this
environment plays a strong role in determining the observations of
these objects.  In this paper, we focus only on the surrounding
environments of AICs.

\subsection{Accretion Induced Collapse Rates}

A number of methods have been used to constrain AIC rates.  Thus far,
no outburst from the accretion induced collapse of a white dwarf has
been observed.  Given that the outburst is expected to be very dim
because shock heating is negligible and the predicted $^{56}$Ni yields
are all low, i.e. $<0.05$\,M$_\odot$ (\cite{fryer99a,kit06,Des07b}),
the lack of observed AICs does not place firm upper limits on the AIC
rate.  With more firm observational predictions and upcoming surveys,
observations will begin to place constraints on the rates.

Theoretical estimates of the rate of AICs are also quite uncertain.  A
number of progenitor scenarios have been proposed, mostly in the
search for the elusive progenitor to type Ia supernovae (see Livio
2001 for a review).  Although single degenerate models exist, the
dominant progenitor of AICs, if we accept the conclusions of
\cite{nomoto91}, comes from double degenerate mergers with a rate of
$\sim$\,10$^{-2}$\,y$^{-1}$ in a Milky-Way sized galaxy.  This result
depends upon a number of assumptions about the accretion evolution in
these binary systems and the true rate of AICs could be many orders of
magnitude lower than this value.  Studies of binary mass
transfer~(\cite{Yoon07,Liv01} and references therein) and white dwarf
accretion (\cite{lund06}) are both becoming more accurate.  As they
are coupled with stellar evolution models of these systems, this rate
estimate for AICs should become more accurate.

Alternatively, one can use observed features of AIC explosions to
constrain the AIC rate.  By comparing observations of nucleosynthetic
yields to explosion models, \cite{fryer99a} argued that the neutron
rich ejecta from an AIC limits their rate to
$\sim$\,10$^{-4}$\,y$^{-1}$.  More recent results, which eject a
smaller fraction of neutron-rich material, may loosen this constraint
by 1 order of magnitude (\cite{kit06,Des07b}), allowing rates as high 
as $\sim$\,10$^{-3}$\,y$^{-1}$.

\subsection{AIC Light Curves and Spectra}

Recall that light curves and spectra are powered by both shock heating
as the ejecta hits its surrounding medium and through the decay of
radioactive elements.  The standard explosion model of \cite{fryer99a}
predicted $\sim$0.05\,M$_\odot$ of $^{56}$Ni ejecta.  Other explosion models
predict even less mass in $^{56}$Ni.  In this case, shock heating will
play an equal, if not dominant, role in the light curve.

\begin{figure}
\plotone{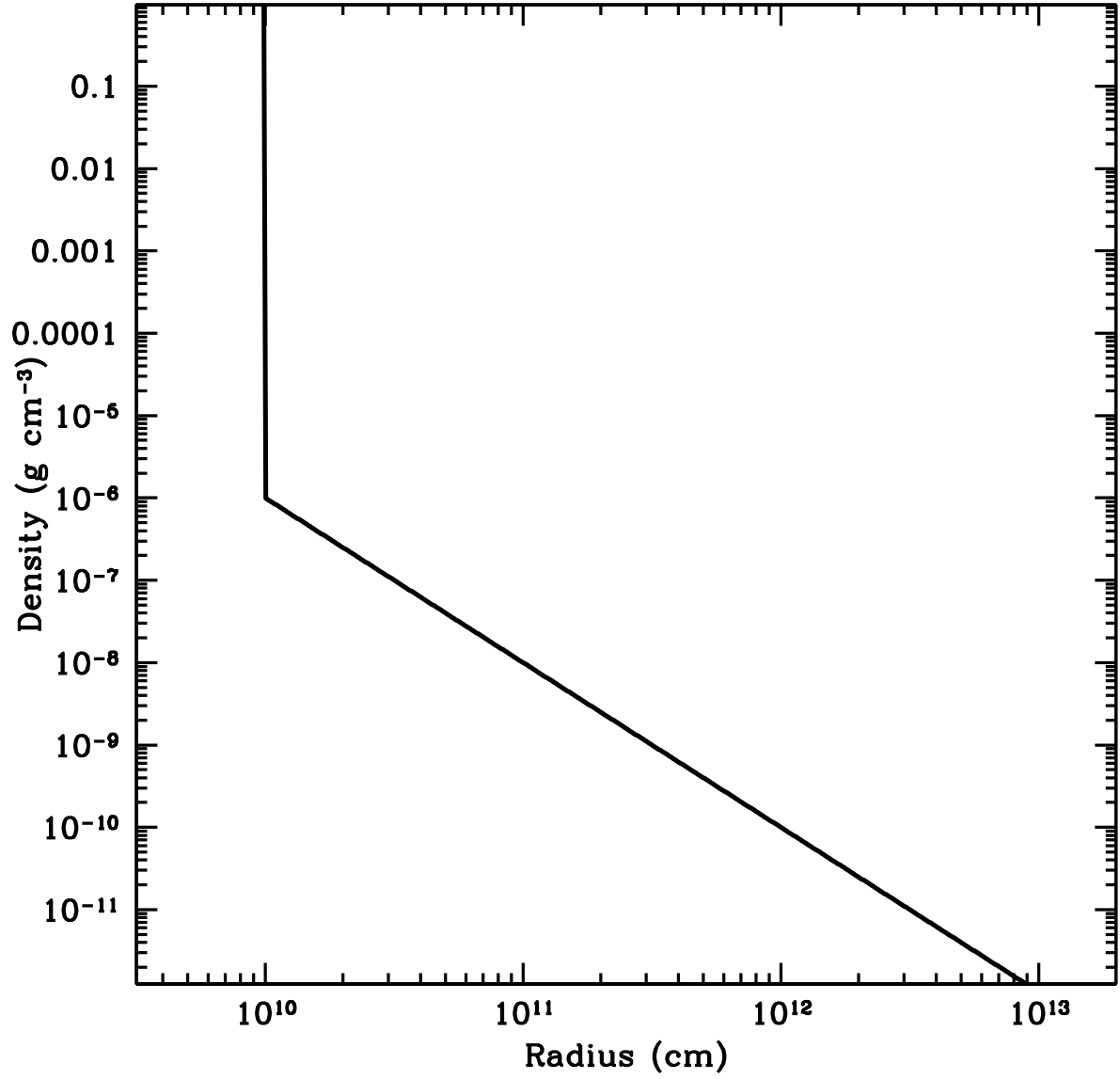}
\caption{Density distribution (density versus radius) for our AIC calculations.}
\label{fig:aicdens}
\end{figure}

For our calculations, we use an explosion from \cite{fryer99a}.  The
total explosion energy for our canonical AIC is $2\times10^{51}$\,ergs.
With the low ejecta mass (0.2\,M$_\odot$), this energy corresponds to
a high average initial velocity of the ejecta ($3 \times
10^{9}$\,cm\,s$^{-1}$).  The composition is also based on the
explosion models of \cite{fryer99a}, with roughly 20\% of the eject in
the form of $^{56}$Ni (0.04\,M$_\odot$).  We construct a second
explosion with 1/10th the amount of mass (hence 1/10th the explosion
energy and 1/10th the $^{56}$Ni yield) to compare to the lower ejecta
models predicted by more recent calculations~\cite{Des07b}.  A summary
of the explosion energy, ejecta mass, and $^{56}$Ni mass is shown in
Table~\ref{table:cond}.  On top of these explosions, we construct a
surrounding environment with a density structure
(Fig.~\ref{fig:aicdens}) based on preliminary binary merger
calculations by Motl et al. (in preparation).

\begin{figure}
\plottwo{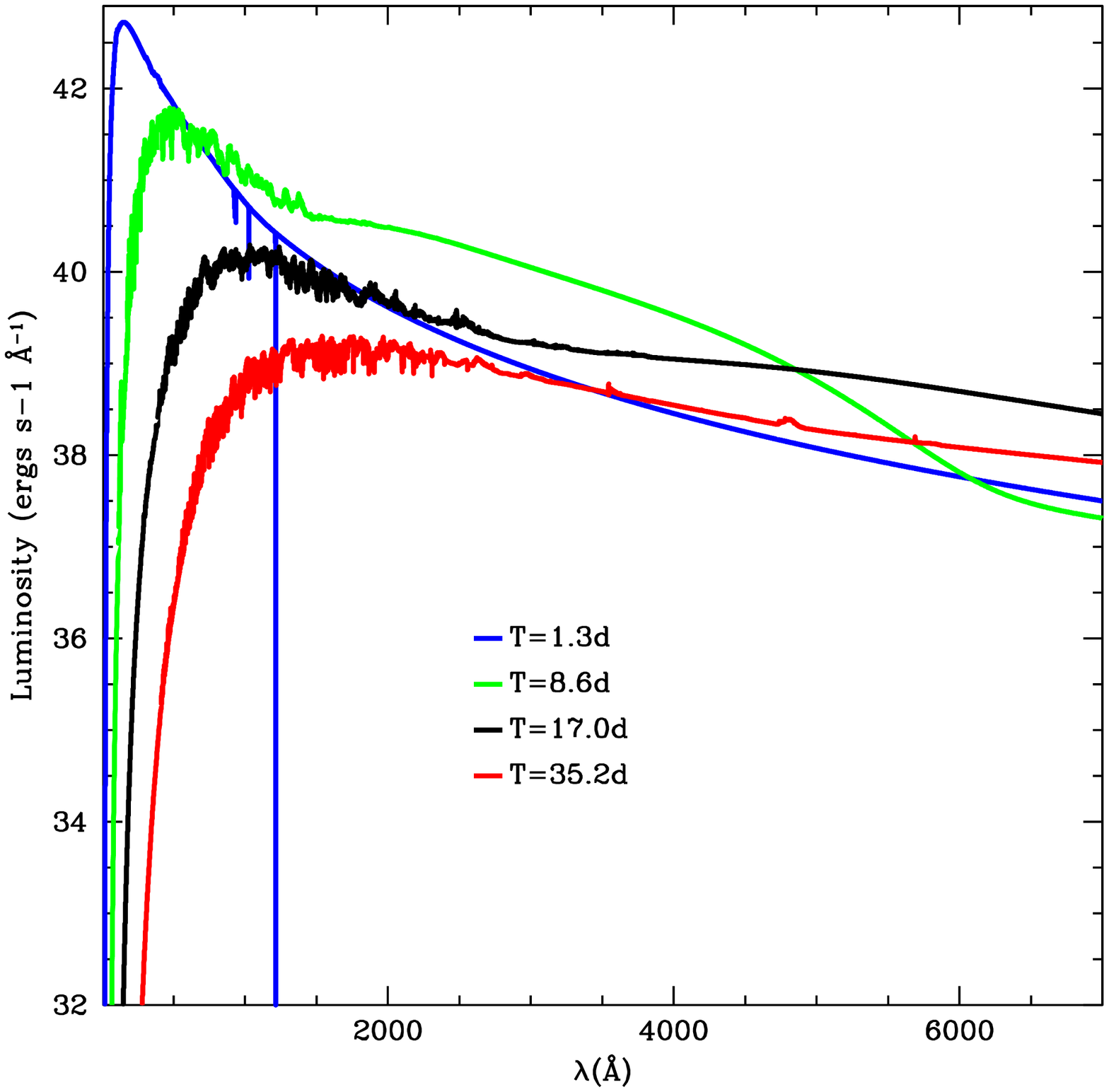}{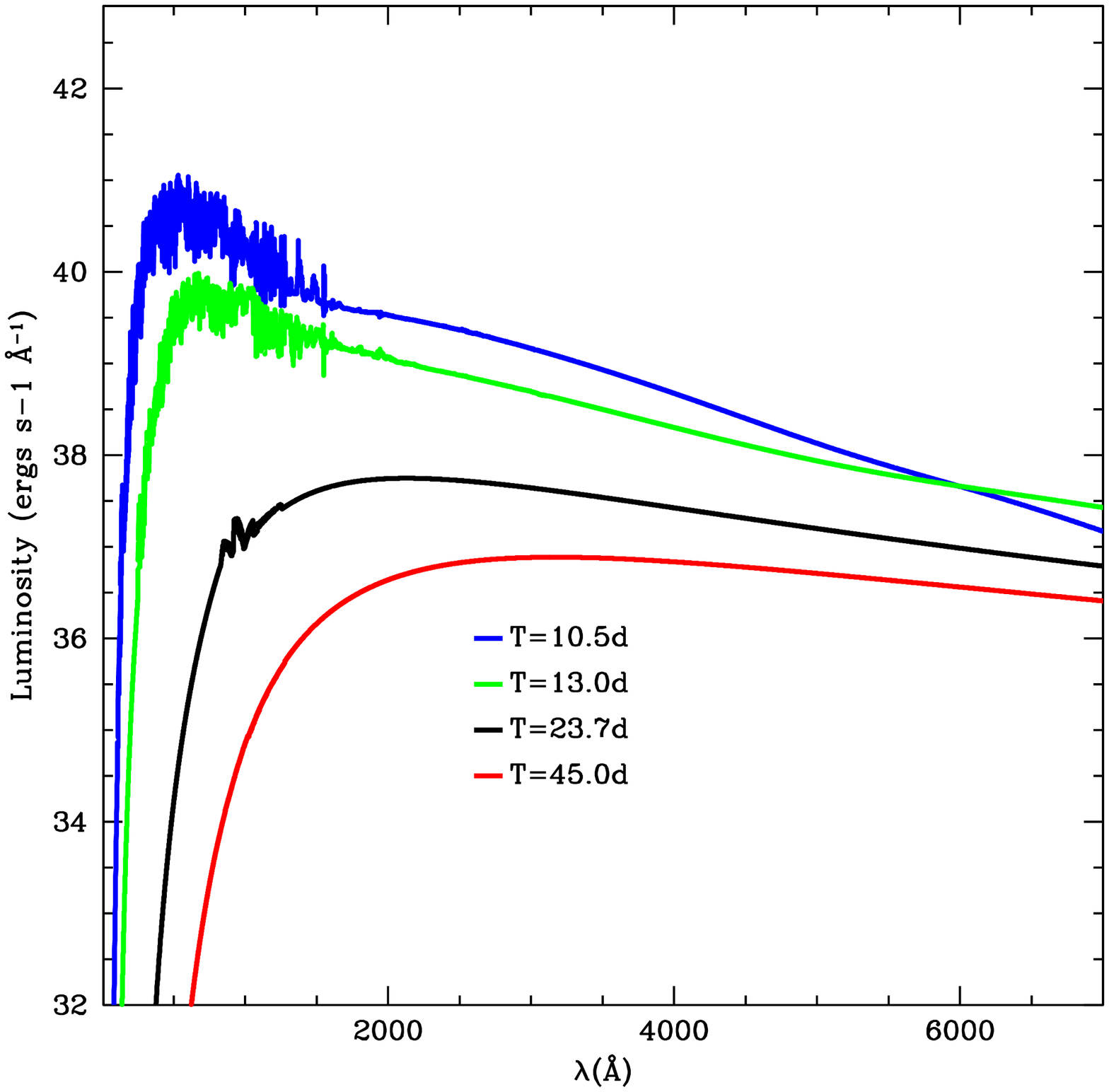}
\epsscale{0.5}
\plotone{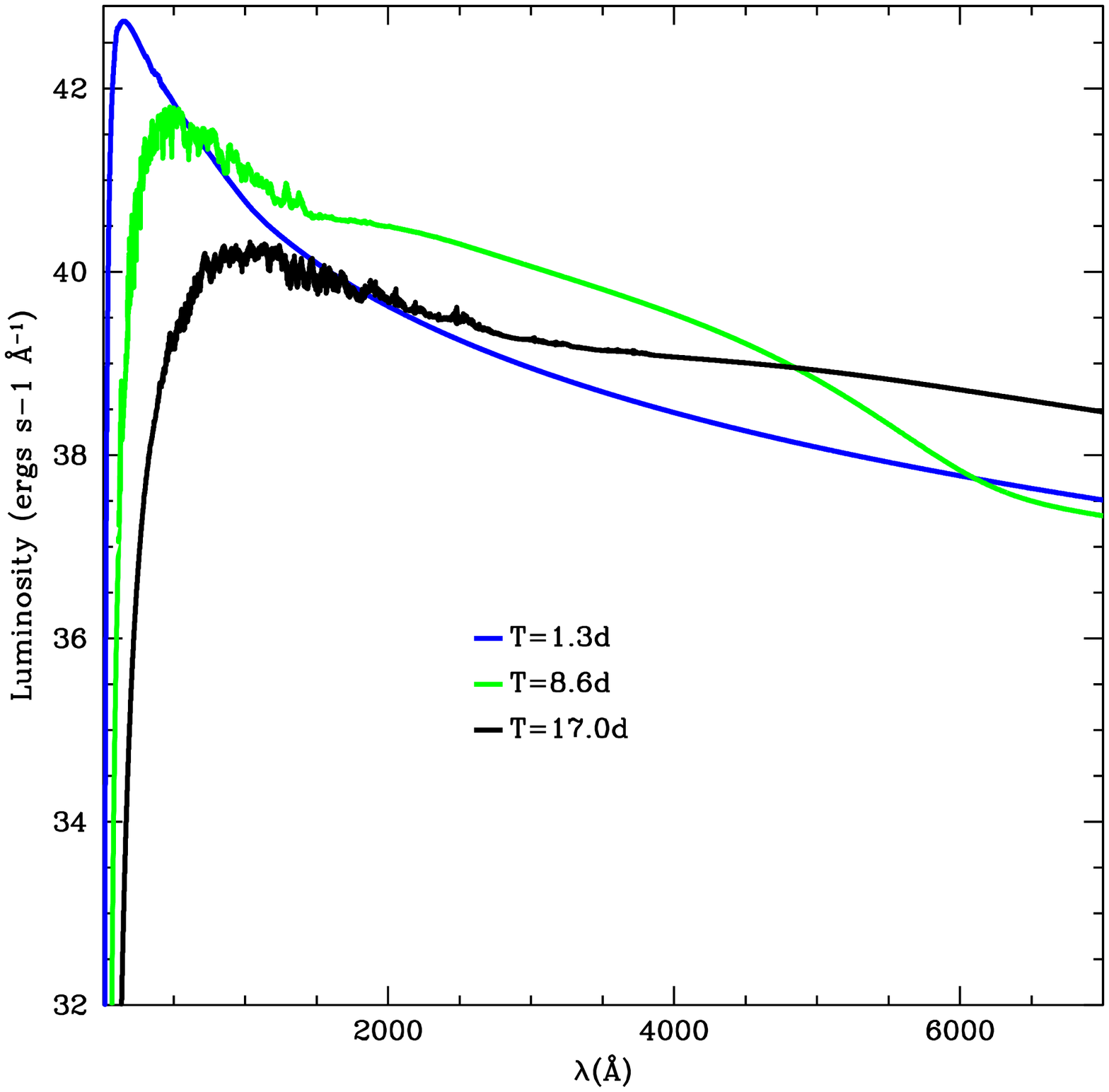}
\epsscale{1.0}
\caption{Snapshots in time of our AIC spectra for 3 models: The top two
plots represent a large
$^{56}$Ni yield of 0.041\,M$_\odot$ (left) and a low $^{56}$Ni yield of
0.0041\,M$_\odot$ (right).  The lower $^{56}$Ni yield coupled with
lower energy produces a much weaker explosion.  Note that the spectra
peak below 1000\,\AA\ (in the X-ray).  The bottom plot represents our large
$^{56}$Ni yield with CO on top of the star instead of helium.  Note
that, for this model, the composition on top of the explosion plays 
very little role in shaping the spectrum.}
\label{fig:aicspec}
\end{figure}

Due to the low envelope mass, the ejecta begin to emit within the
first day of the explosion (Fig.~\ref{fig:aicspec}).  At this time,
the ejecta is still hot, ionizing the material above it, leading to
very few lines.  Even when we place a CO atmosphere on top of the
white dwarf, at early times there are very few lines due to the high
peak temperature.  As the ejecta expands, it cools and line features
appear.  But there are no strong identifying (specific to AICs) 
features in the spectra.

\begin{figure}
\plottwo{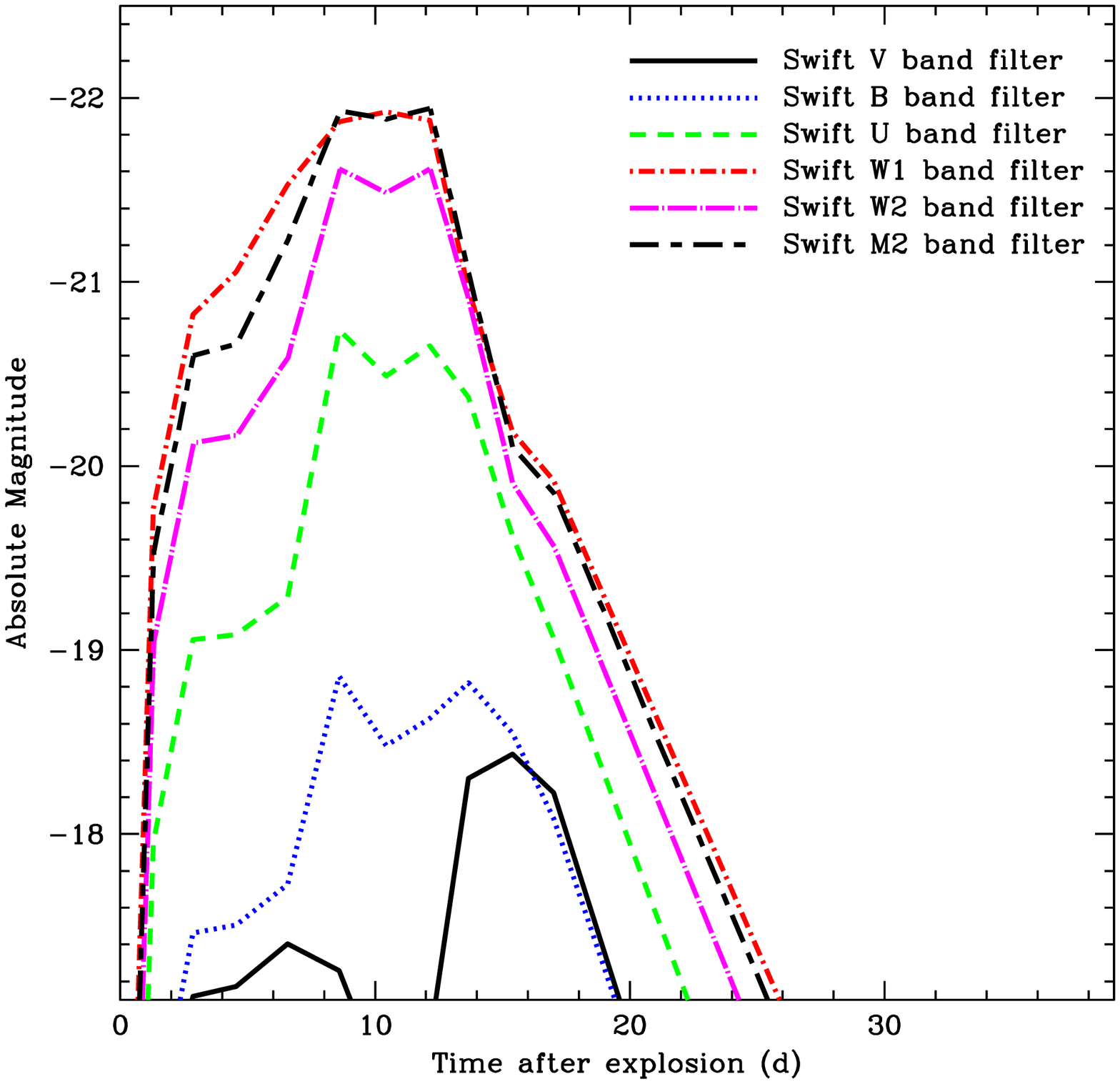}{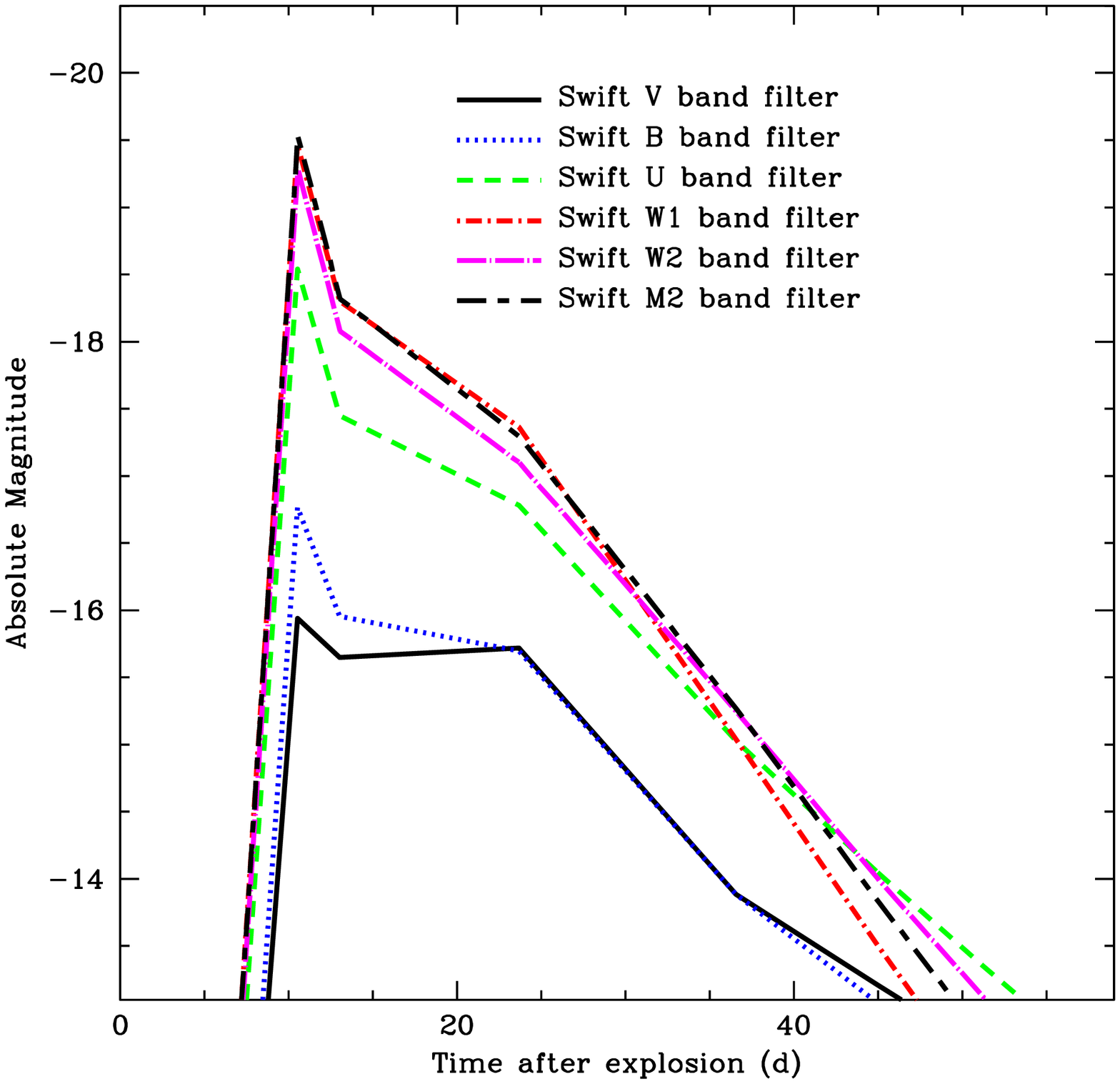}
\caption{Light curves (absolute magnitudes as a function of time) for
5 different bands based on the Swift filters.  The peak in the X-ray
for these explosions means that the higher-energy bands are brightest.
Although the visible magnitudes are dimmer than normal Type Ia
supernovae, these supernovae are much brighter than normal Ia
supernovae in the high-energy Swift W1 and W2 bands.  The left panel 
shows our high-mass ejecta run with 0.041\,M$_\odot$ of $^{56}$Ni 
ejecta.  The right panel shows the light curves for our low-mass 
ejecta run (0.0041\,M$_\odot$ of $^{56}$Ni ejected).}
\label{fig:lc-aic}
\end{figure}

The light curves in V and B bands peak at roughly 15\,d with peak
absolute magnitudes of -18.5 to -19 magnitudes (close to that of
supernovae) for our standard model (Fig.~\ref{fig:lc-aic}).  The drop
is fairly rapid, and by 30d, the absolute magnitude for both these
bands is below -17.  The peak in the U and Swift UV bands is bright
(as we might expect from the high effective temperatures of the
spectra).  

For our lower yield model, the combined lower energy and lower mass of
$^{56}$Ni ejected lowers the AIC emission.  The peak absolute V and B
magnitudes of our low-density run do not exceed -16 and by 40\,d are
below -14.  At early times, there are a number of lines in the UV, but
by 20\,d, the low density and high temperature of this model ionizes
most of the material and the spectra are fairly featureless.

\subsection{Neutrinos and Gravitational Waves from AICs}

The collapse and bounce of an AIC is very similar to that of a normal
supernova (see Fryer \& New 2003 for a review).  As such, both the
neutrinos and GW emission should be similar to that of core collapse.
The primary difference is that the explosion is likely to happen
quickly and there is unlikely to be much, if any, material falling
back onto the newly formed neutron star.  For neutrinos, this means
that the explosion is clean, allowing a clear view of a cooling
neutron star.  For gravitational waves, there will be no strong signal
from convective instabilities above the proto-neutron star.  But there
is a possibility that AICs will have high angular momenta at collapse.
As such, the AIC scenario is a leading candidate among stellar
collapse to form bar-mode (and related) instabilities.

\section{Fallback Supernovae}
\label{sec:fallback}

Standard core-collapse supernovae are the explosions produced when the
iron cores of stars more massive than 8--10\,M$_\odot$ collapse to
form neutron stars.  The potential energy released in this collapse
drives an explosion.  But not all stellar collapses form strong
supernova explosions.  The explosion launched in the stellar core
moves out through the star and decelerates as it pushes out the rest
of the star.  For some of the initial exploding material, this
deceleration drops the explosion energy below the escape energy from
the core.  This material ultimately falls back onto the neutron star.

\cite{Fry99} argued that although this ``fallback'' is only a few
tenths of a solar mass for 15\,M$_\odot$ stars, it might be several
solar masses for 25\,M$_\odot$ stars.  Based on their understanding of
fallback, Fryer \& Kalogera (2001) argued for a range of neutron star
and black hole masses.  One of the successes of the current supernova
mechanism is its prediction of fallback and a broad range of remnant
masses.  But fallback also has implications for supernova light
curves.  $^{56}$Ni is produced in the innermost ejecta.  This ejecta
is the first material to fall back and if the fallback is extensive,
very little $^{56}$Ni will be ejected to power the supernova emission.
In addition, fallback tends to occur in weaker explosions, reducing
the emission energy from shocks as well.  So fallback supernovae will
have a range of peak emission, from energies as strong as classic
supernovae when little fallback occurs down to unobservable whimpers
when the fallback is extensive.

The fate of the core changes if fallback is so large that it pushes 
the neutron star beyond the maximum neutron star mass.  These systems 
collapse to form a black hole.  For this paper, we will focus on the 
emission of black-hole-forming, fallback supernovae.  

\subsection{Fallback Rates}

In current simulations, the energy produced in the convective engine
decreases for stars more massive than 20\,M$_\odot$ while the binding
energy of the star increases dramatically at roughly this same mass
point.  Including large errors in the explosion energies, \cite{FK01}
were able to pinpoint the transition mass from neutron star and black
hole formation to stars with initial masses within the
18--23\,M$_\odot$ range.  \cite{FK01} found that, within
the uncertainties, the formation rate of black holes in supernova
explosions was somewhere between 10--40\% that of the total supernova
rate.  The largest uncertainty in this estimate is the initial mass
function.  Winds can allow the formation of neutron stars by more
massive, solar metallicity stars (above $\sim 60$\,M$_\odot$), but this
does not affect the rate significantly.

\subsection{Fallback Spectra and Light Curves}

\begin{figure}
\plotone{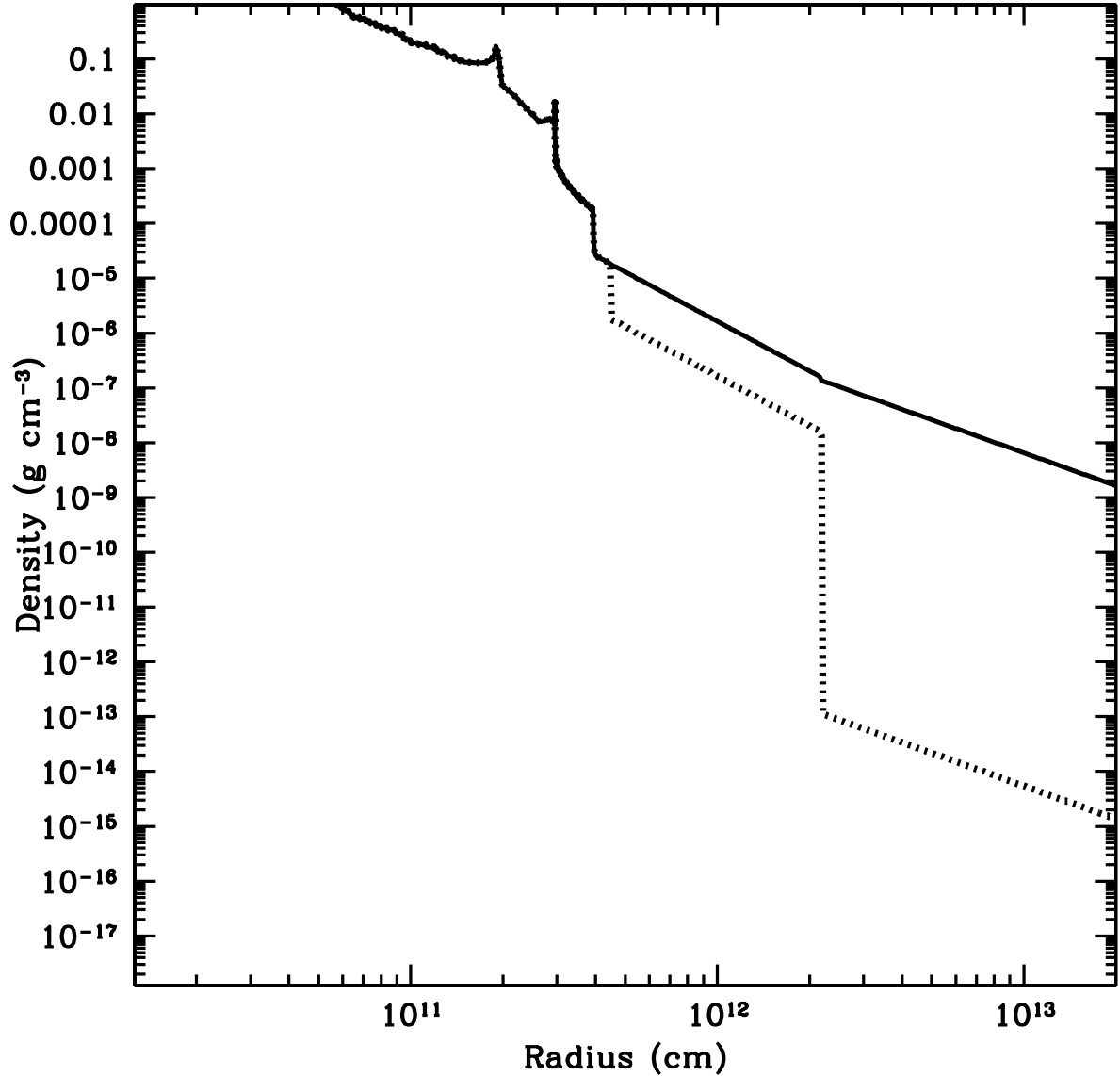}
\caption{Density profiles for our fallback supernova calculations.  In
  one case, we use a dense medium corresponding to a late-time binary
  mass ejection scenario (solid line).  In the other case (dotted
  line), we assume a Wolf-Rayet wind medium
  ($\dot{M}=10^{-5}\,M_\odot\,y^{-1}$).}
\label{fig:fbdens}
\end{figure}

For our calculations, we use a 40\,M$_\odot$ binary progenitor for Cas
A (Fragos et al. 2008).  In this star, we drive a
$2\times10^{51}$\,erg explosion.  The binding energy of this star is
much greater than $2\times10^{51}$\,ergs and the final remnant mass
after fallback is 4.5\,M$_\odot$.  With this much fallback, very
little $^{56}$Ni is ejected: $< 2 \times 10^{-13}$\,M$_\odot$.  A summary of
the explosion energy, ejecta mass, and $^{56}$Ni mass is shown in
Table~\ref{table:cond}.  On top of this explosion, we construct two
surrounding environments with density structures
(Fig.~\ref{fig:fbdens}): one based on binary mass ejection
(100\,km\,s$^{-1}$ velocity, $\dot{M}=1$\,M$_\odot$\,y$^{-1}$) and
one with a small atmosphere ($< 0.00054\,M_\odot$) topped by a wind
profile (1000\,km\,s$^{-1}$ velocity,
$\dot{M}=10^{-5}$\,M$_\odot$\,y$^{-1}$).

\begin{figure}
\plottwo{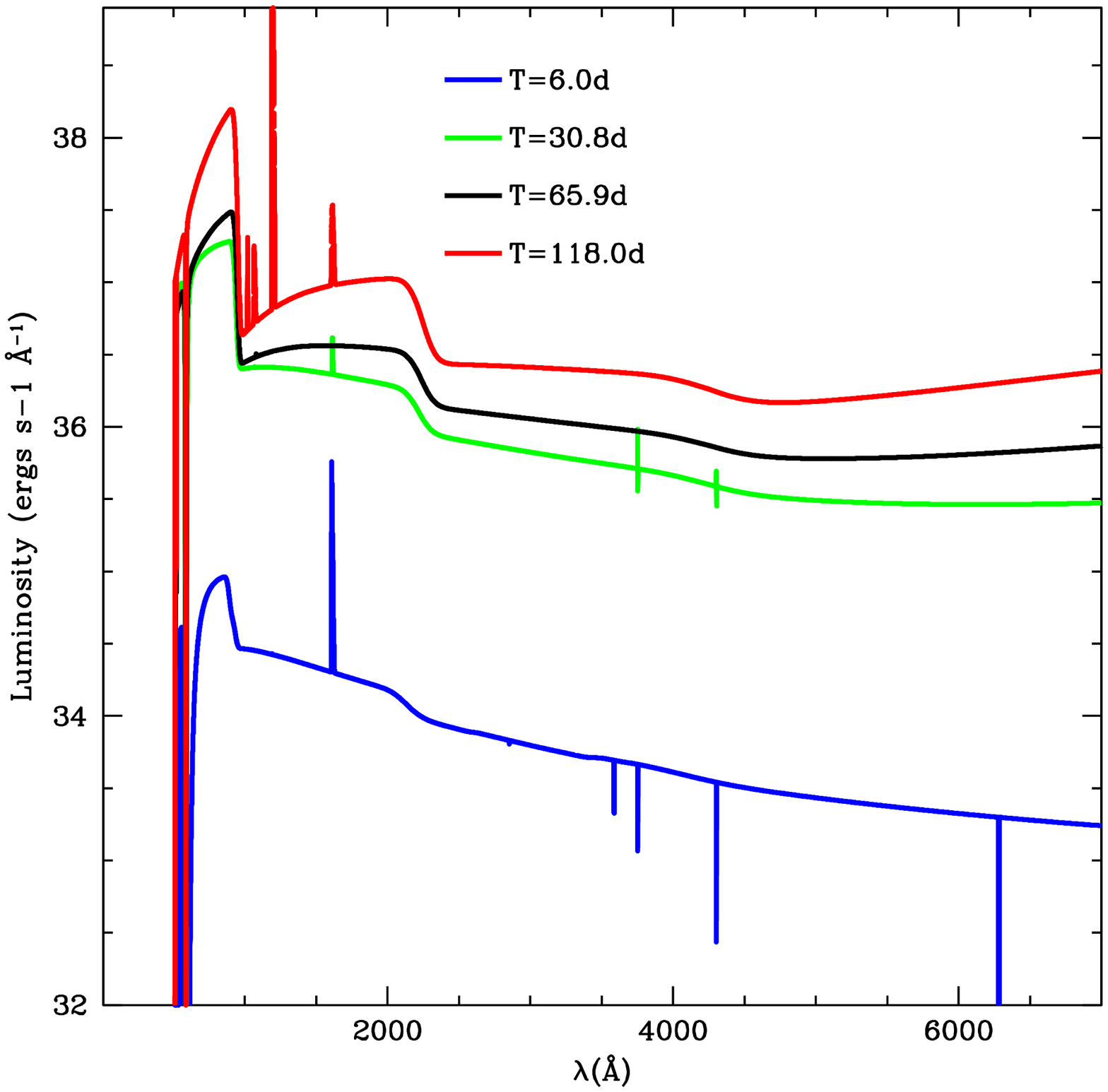}{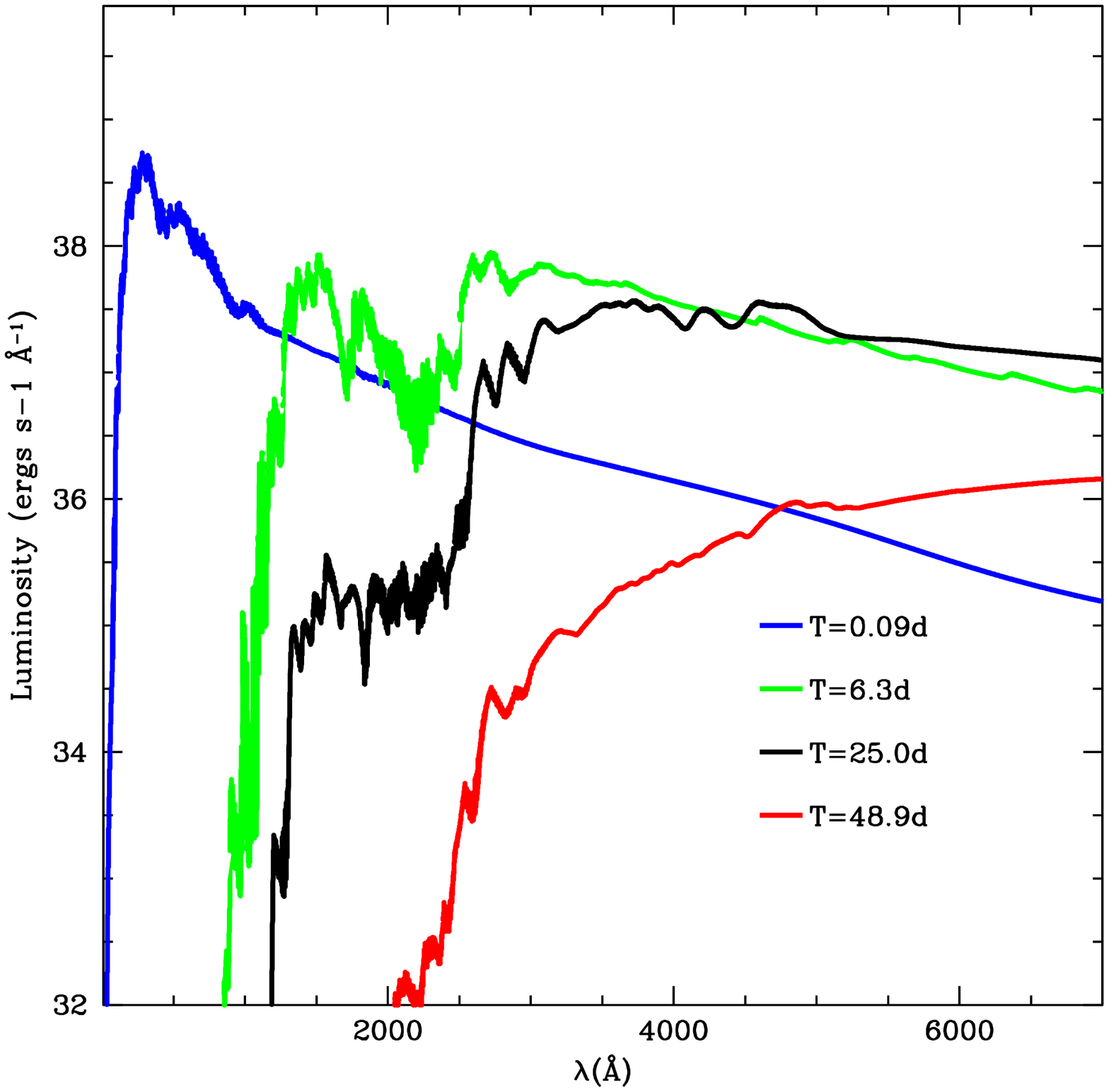}
\plottwo{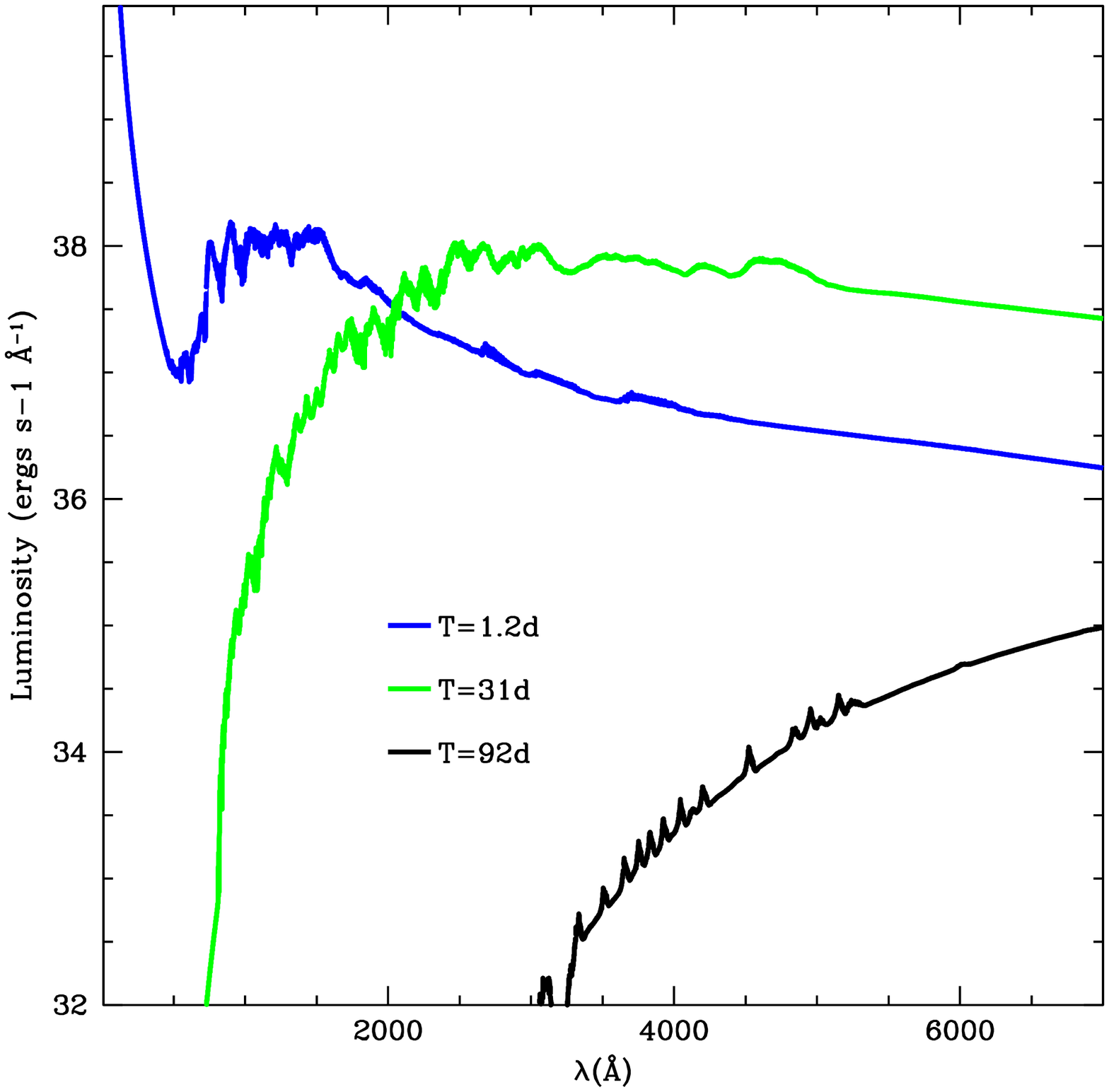}{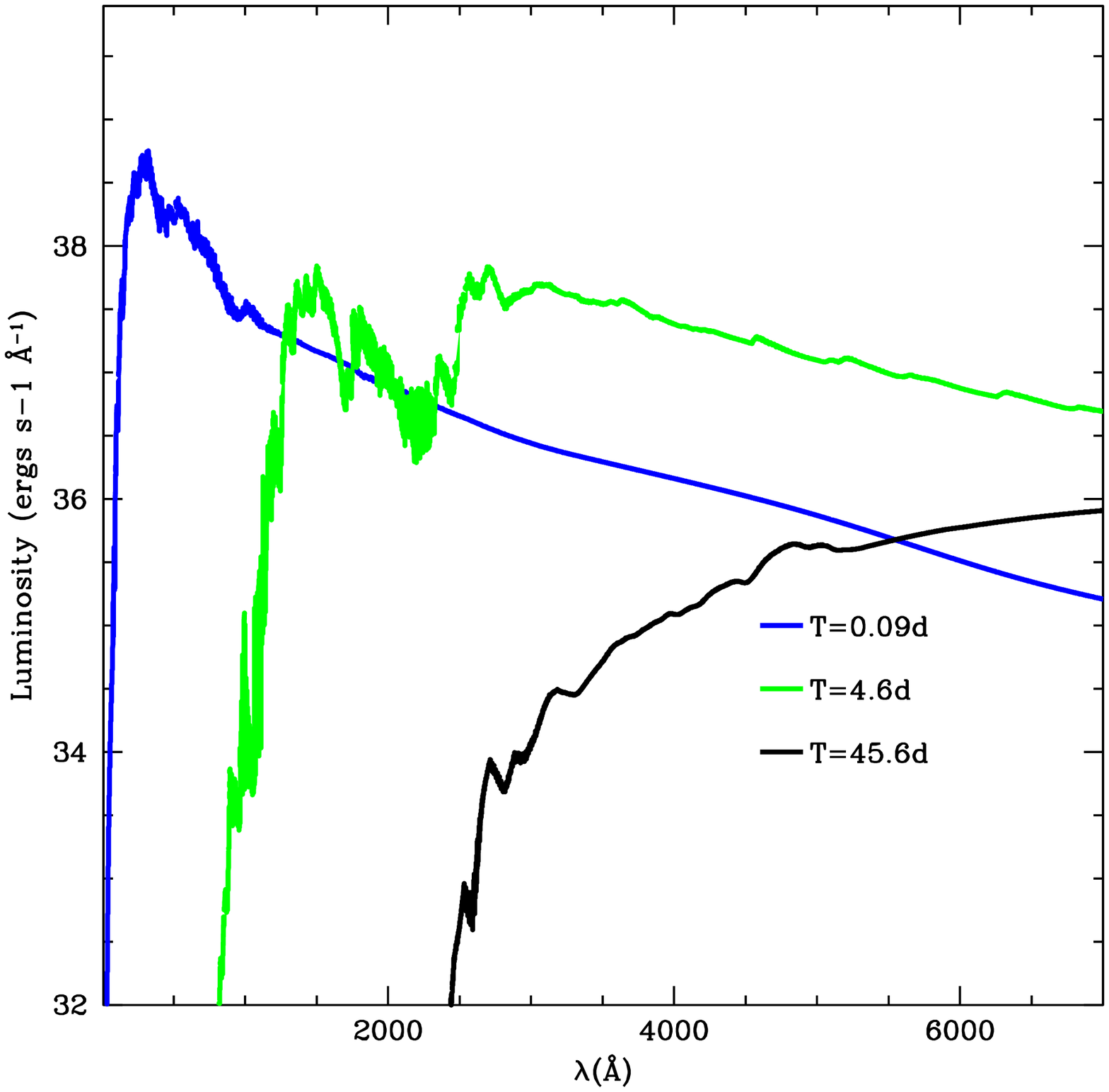}
\caption{Spectra from our extremely dense and dense surrounding medium
models of supernova fallback.  In the more dense case, the outburst is
still peaking 100\,d after the launch of the explosion and strong
emission/absorption lines are evident. The lines, dominated by
material ahead of the ejecta, are narrow.  In the lower density case,
the peak occurs much sooner and lines are part of the ejected material
and hence are broader.  The bottom, left plot shows the lower density case 
using 5 groups for the radiation transport.  The additional groups 
lead to a slightly different temperature profile that can change 
the spectra dramatically.  But the peak fluxes are not altered 
significantly.  The bottom right plot shows the results using 
twice the coarse-bin resolution and 10 times the fine (AMR) 
resolution.  The spectra are nearly identical to its comparable 
low resolution case (dense surrounding medium).}
\label{fig:fbspec}
\end{figure}

\begin{figure}
\plottwo{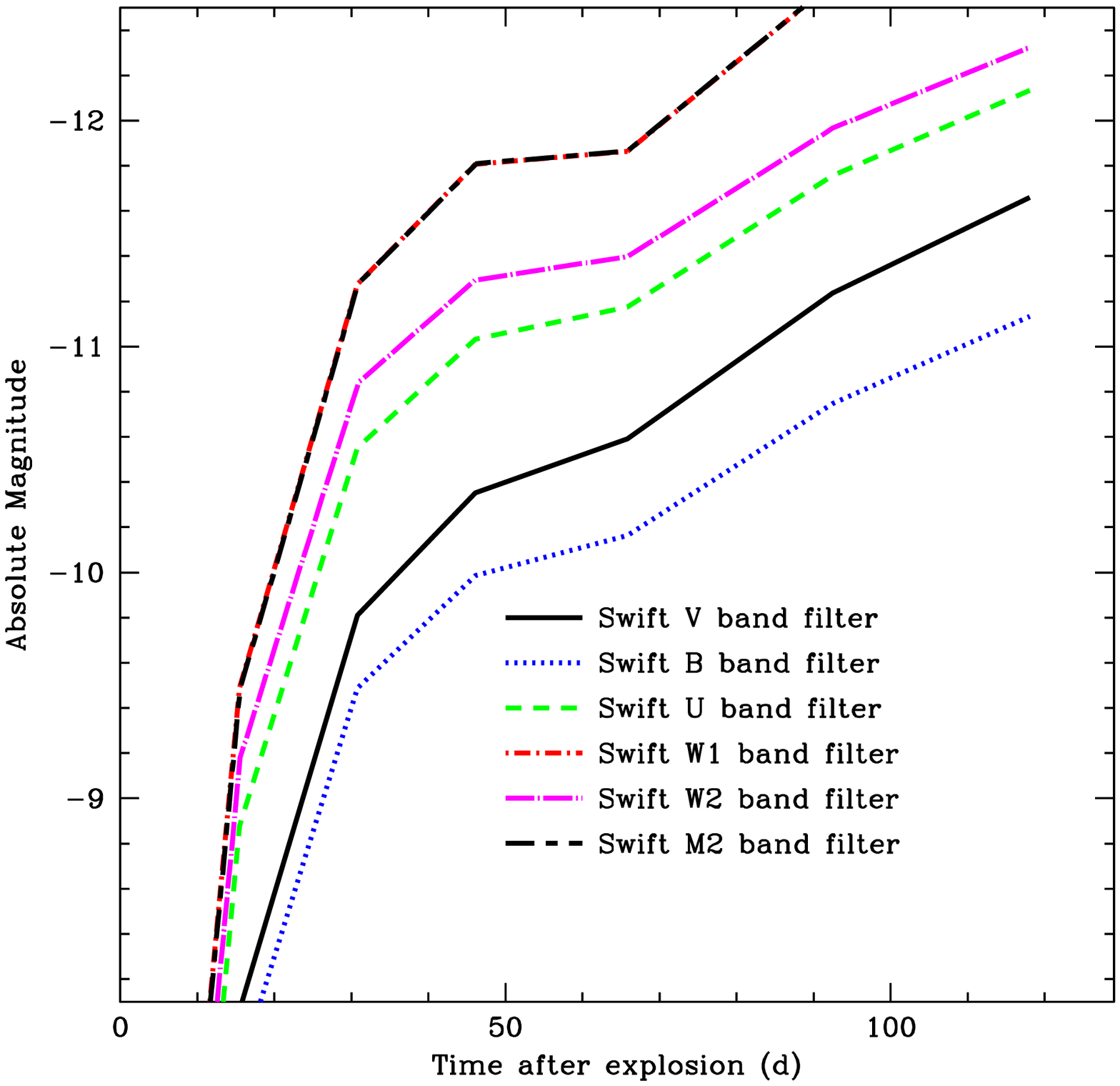}{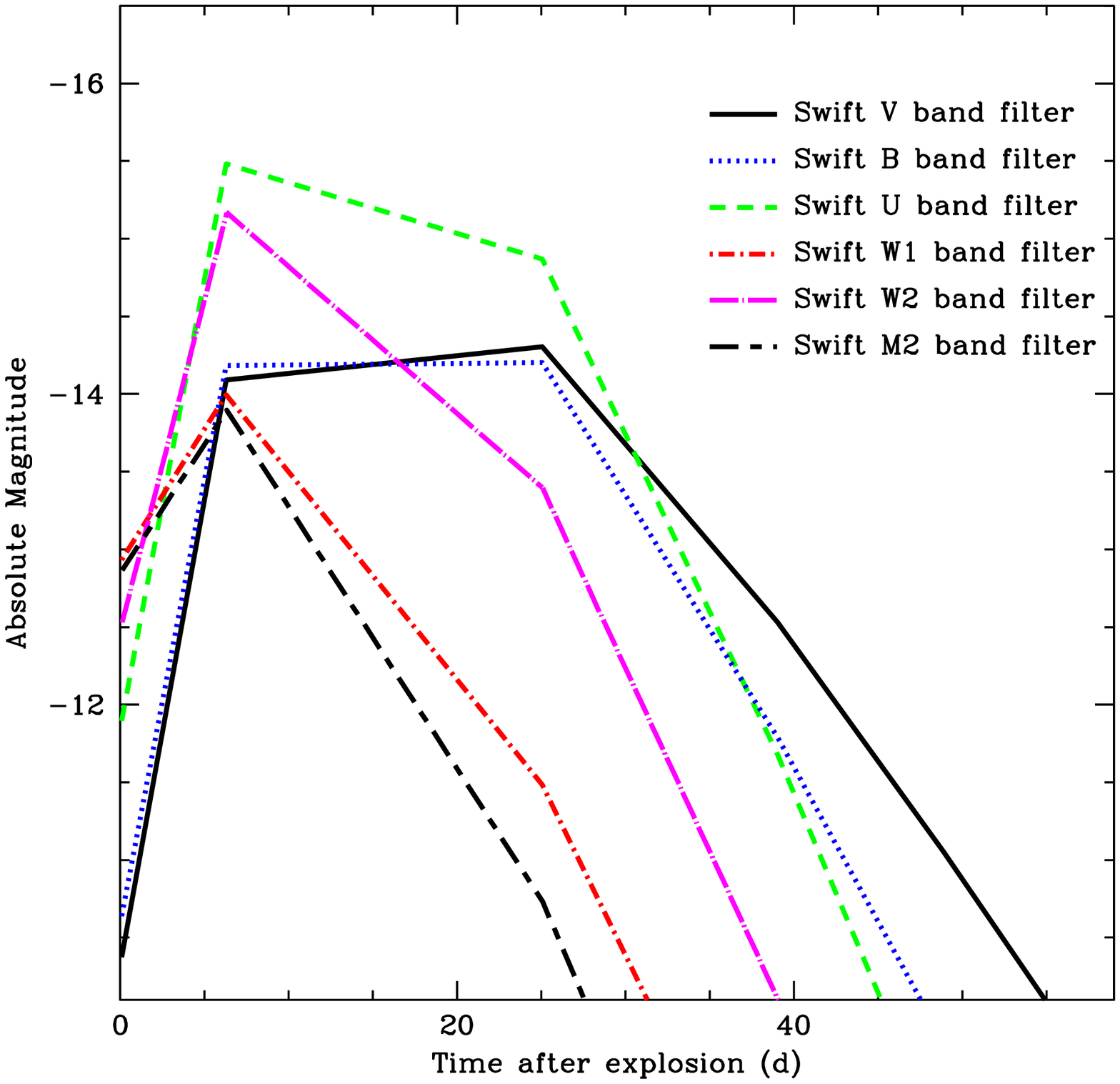}
\caption{In the extremely dense surrounding 
medium model (left panel), the light curve is still rising after 
100\,d.  In the less dense case (right panel), the light curve peaks 
after 15\,d, but is 3 magnitudes brighter than the 
projected peak in our dense model.}
\label{fig:lc-fb}
\end{figure}

The large mass in the binary mass ejection case, coupled to the weak
explosion energy, delays shock breakout until after the ejecta loses
much of its energy (Figs.~\ref{fig:fbspec},\ref{fig:lc-fb}).  In this
simulation, the explosion has yet to peak, even 100\,d after the
explosion.  But it will peak at very low V, B magnitudes (below
absolute magnitudes of -13).  This is an extreme case, where the
environment is very dense out to $10^{16}$\,cm due to a binary mass
ejection just prior to collapse.

More likely, the mass ejection phase is followed by a Wolf-Rayet wind
phase.  Even with this lower-density surrounding medium, the low
ejecta velocity coupled to its low $^{56}$Ni ejecta produces a very
weak explosion with peak V, B absolute magnitudes of -15
(Figs.~\ref{fig:fbspec},\ref{fig:lc-fb}).  The lower density means the peak
emission occurs quickly ($\sim$10\,d) and the V-band absolute
magnitude drops below -12 at about 45\,d.  

To test our single group approximation, we modeled a 5-group transport
calculation for our diffuse case (Fig.~\ref{fig:fbspec}).  Although
many of the lines are similar, the spectral fluxes can be very
different.  Although the peak in the light-curve doesn't change
dramatically in the V-band, the UV light-curve is very different.
Ultimately, many group calculations will be required to model detailed
spectral light curves.

To test our resolution, we completed 1 run with twice the coarse-bin
resolution and 10 times the effective (AMR) spatial resolution
(Fig.~\ref{fig:fbspec}).  The spectra from this simulation is nearly
identical to our standard runs.

\begin{figure}
\plottwo{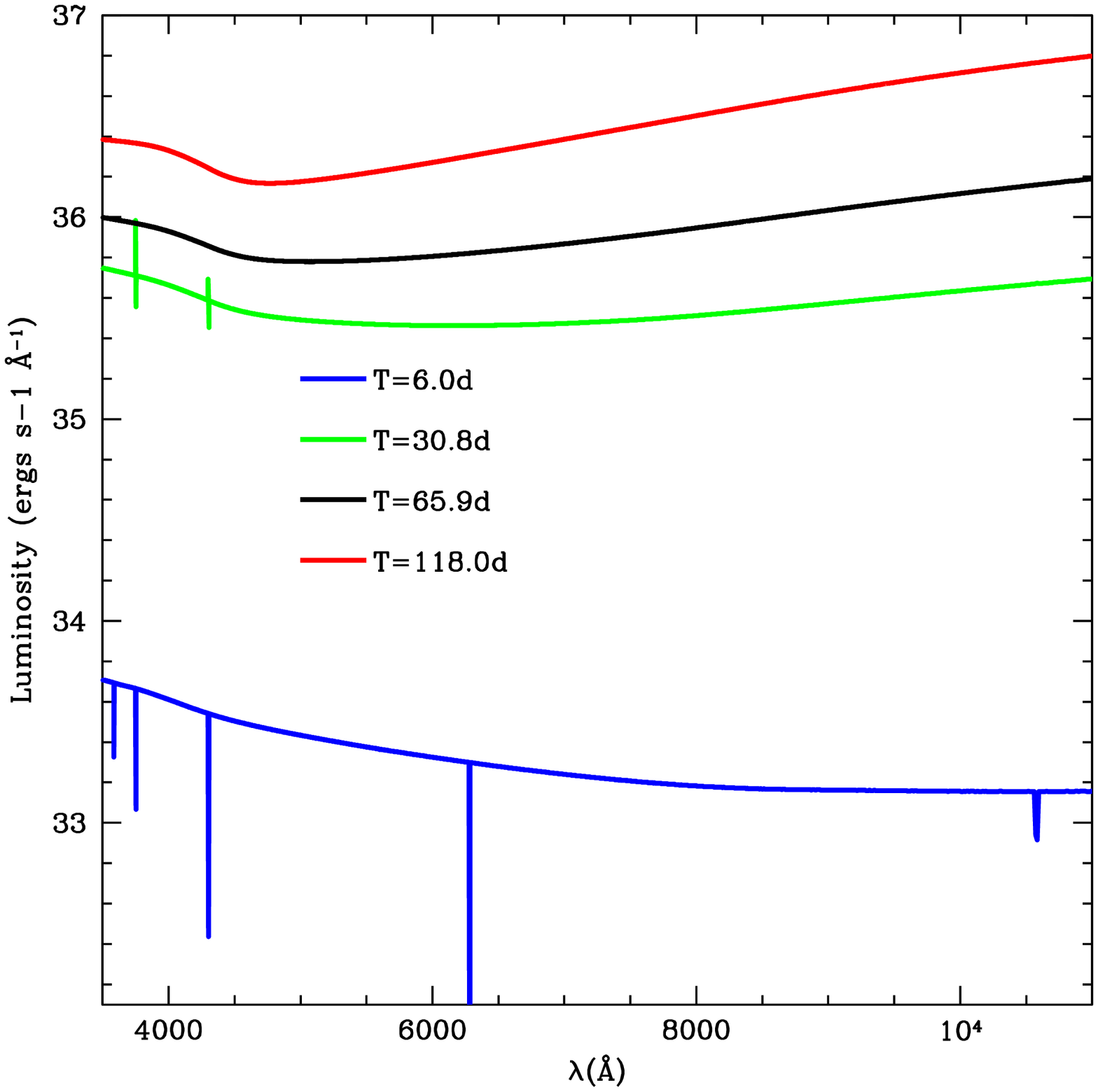}{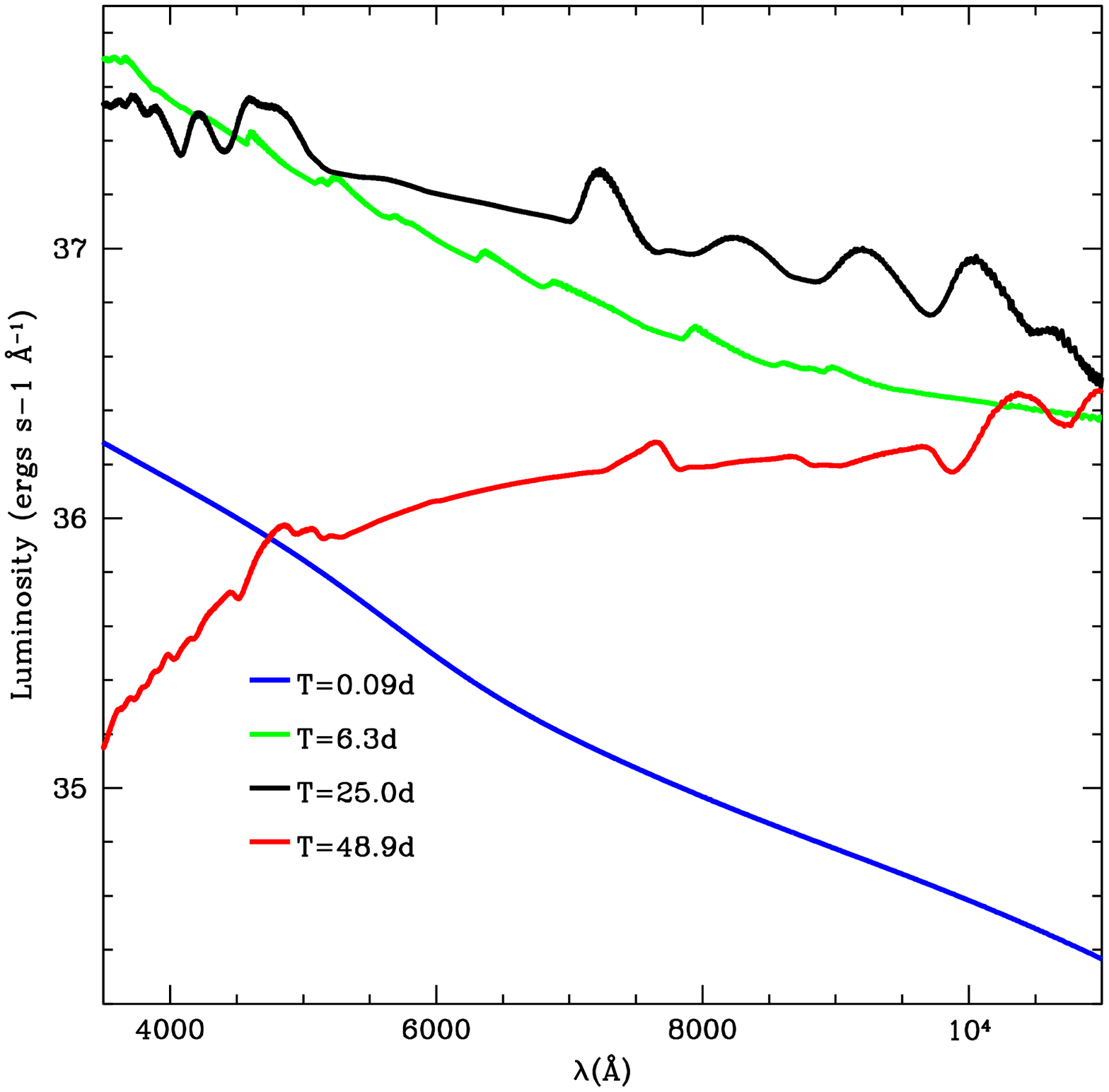}
\caption{Optical and IR spectra of our high and low density fallback
models at the same time snapshots as figure~\ref{fig:fbspec}.  
In the dense model, the narrow lines are produced by material just 
being heated by the radiation front.  First only in absorption, 
as the material is heated, we also see emission lines.  In the 
less dense model, we observe the broad lines from the 
shocked ejecta itself.}
\label{fig:fbopspec}
\end{figure}

Detailed spectra might also help to give us a better understanding 
of the surrounding medium.  Figure~\ref{fig:fbopspec} shows the 
optical/IR spectra for our two fallback models for the same 
snapshots in time of figure~\ref{fig:fbspec}.  Note that we 
assume our atomic levels are in local thermodynamic equilibrium 
with the radiation front.  Especially for material ahead of our 
shock front (which is not in thermodynamic equilibrium), we 
overestimate the level of ionization, producing fewer lines than 
what may be observed.  Within the shocked material, local 
thermodynamic equilibrium is a better assumption, and our 
broad lines representing this shocked ejecta are fairly accurate 
and provide an ideal probe of the explosion itself.

\subsection{Neutrinos and Gravitational Waves from Fallback}

The collapse and bounce phases of stellar collapse with considerable
fallback is similar to normal supernovae.  Explosions with
considerable fallback are weaker explosions.  In general, these
explosions have longer delays between bounce and explosion.  As such,
the convective timescale is longer, producing a longer boiling phase
prior to explosion.  After the explosion, fallback accretion adds mass
to the proto-neutron star, possibly causing it to collapse to form a
black hole.  These engines produce neutrino light curves that are much
broader than normal supernovae.  The total emission will be more than
a factor of 2 higher than normal supernovae, primarily in an extended
convective phase (during the first second) and a higher neutrino flux
in the first 10\,s from fallback \citep{Fry09}.  Observations of 
this extended emission will constrain our understanding 
of the supernova explosion mechanism and the nature of fallback.

The extended convective phase may also develop low-mode instabilities,
leading to stronger GW emission, especially through asymmetric
neutrino emission \citep{Kot09}.  If the proto-neutron star collapses
to form a black hole, black hole ringing and related instabilities may
occur, producing another source of GWs (see Fryer \& New 2003).

\section{.Ia Supernovae}
\label{sec:type.Ia}

\cite{Bil07} have argued that faint supernova-like outbursts
can occur in helium flashes of accreting material in AM Canum
Venaticorum (AM CVn) binaries.  In these binary systems, a C/O white
dwarf is accreting from its smaller He white dwarf companion, slowly
whittling away the mass of the He white dwarf.  At high accretion
rates, the helium accretes onto the C/O white dwarf and burns stably.
The system evolves, widening the orbit, and causing the accretion rate
to decrease.  At sufficiently low accretion rates ($\dot{M}<2 \times
10^{-6}$\,M$_\odot$\,y$^{-1}$), the burning can be unstable.  The
accreting white dwarf can go through a series of flashes, each
increasing the entropy of the system, leading to larger ignition
masses.  Typically, the last ``flash'' will result in the largest 
explosion and it is this explosion that \cite{Bil07}
focus on as a potential transient observation.

\subsection{.Ia Supernova Rates}

Using the local Galactic density of AM CVn's and assuming every 
AM CVn gives one explosive last flash, \cite{Bil07}
argued that the .Ia supernovae should occur at a rate 
of $6.67 \times 10^{-5}-0.0002$ per year in an E/SO galaxy.

\subsection{.Ia Supernova Spectra and Light Curves}

For our calculations, we use a FLASH~\citep{Fryx00} simulation of a
Type .Ia outburst.  This model produced a $7\times10^{51}$\,erg
explosion with 0.1\,M$_\odot$ of ejecta.  This is our fastest
explosion, with mean velocities around 80,000\,km\,s$^{-1}$.  The
total amount of $^{56}$Ni ejected is 0.014\,M$_\odot$.

\begin{figure}
\plotone{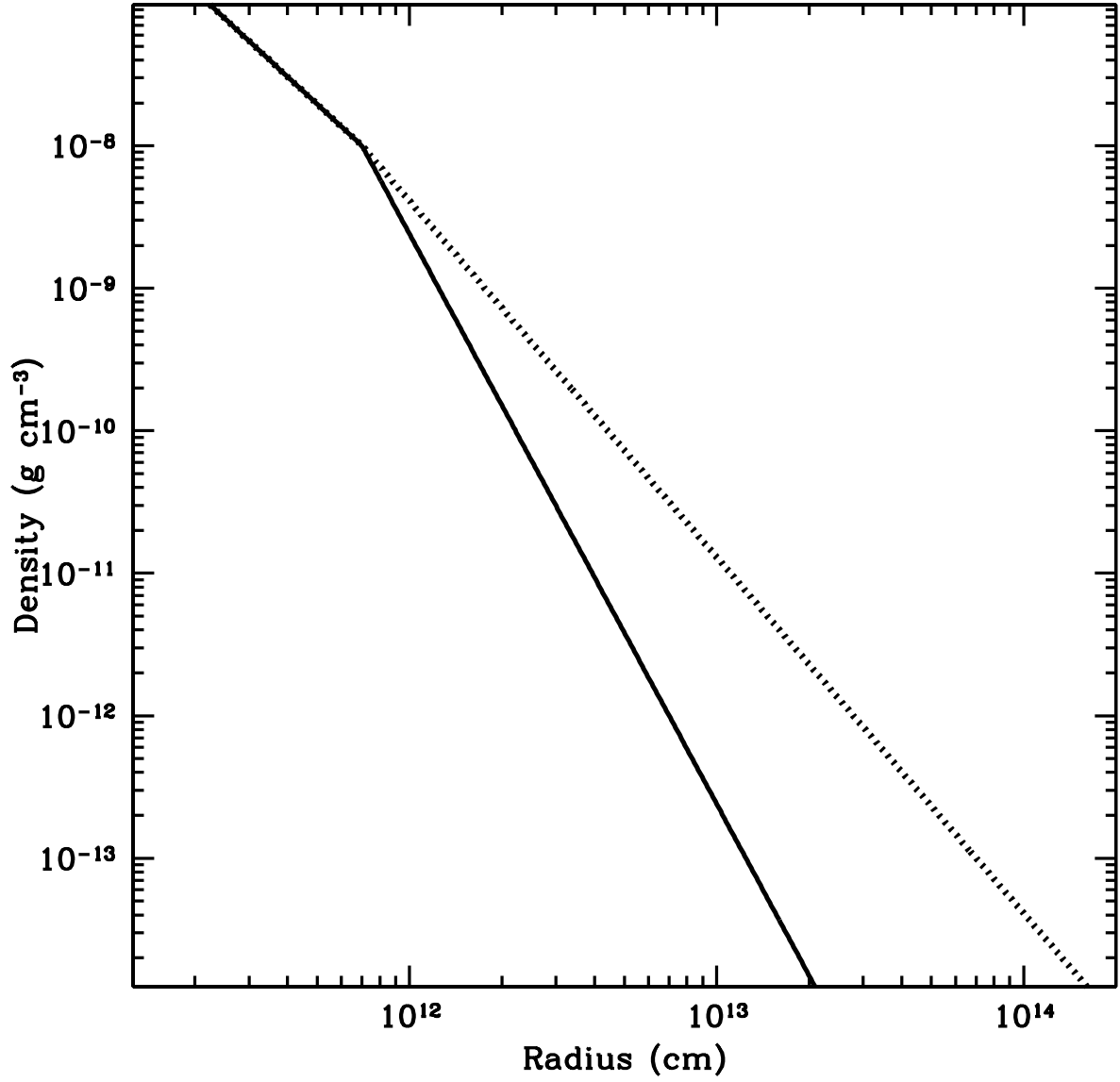}
\caption{Two different density profiles used for our .Ia supernova
models.  The density profiles are fits to binary merger calculations:
one fit along the orbital plane (dotted line) and the other along the
orbital axis (solid curve). }
\label{fig:.Iadens}
\end{figure}

On top of this explosion, we construct two surrounding environments
with density structures (Fig.~\ref{fig:.Iadens}) based on binary
accretion simulations (Motl et al. in preparation): one using the
density profile along the binary orbital plane (higher density) and one
using the profile along the orbital axis (lower density).

\begin{figure}
\plottwo{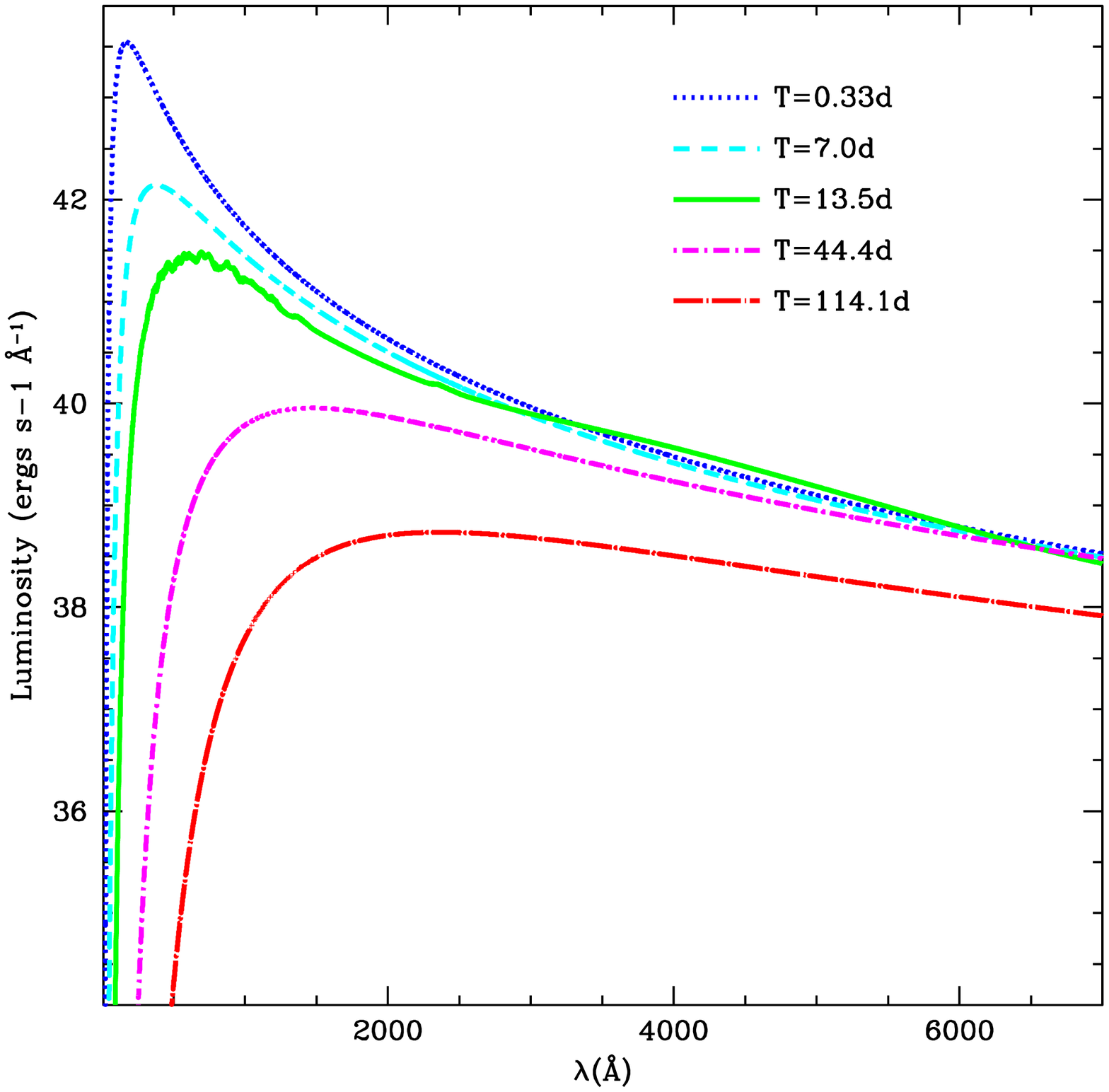}{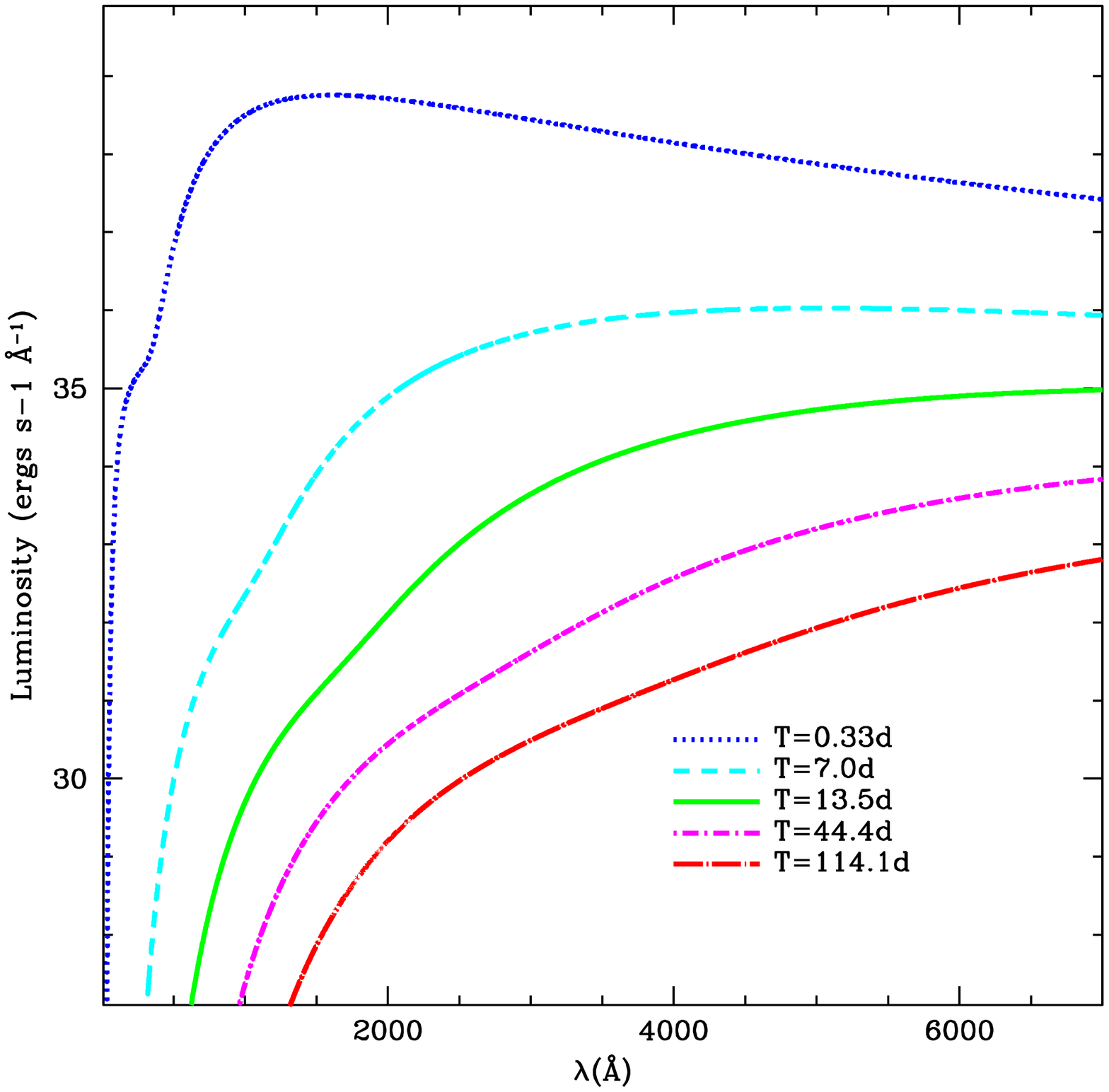}
\caption{Spectra from our .Ia models.  This fast 
shock in this models ionizes most of the material, 
producing line-free spectra.  This strong shock 
also produces spectra that peak in the X-ray.}
\label{fig:spec-.Ia}
\end{figure}

\begin{figure}
\plottwo{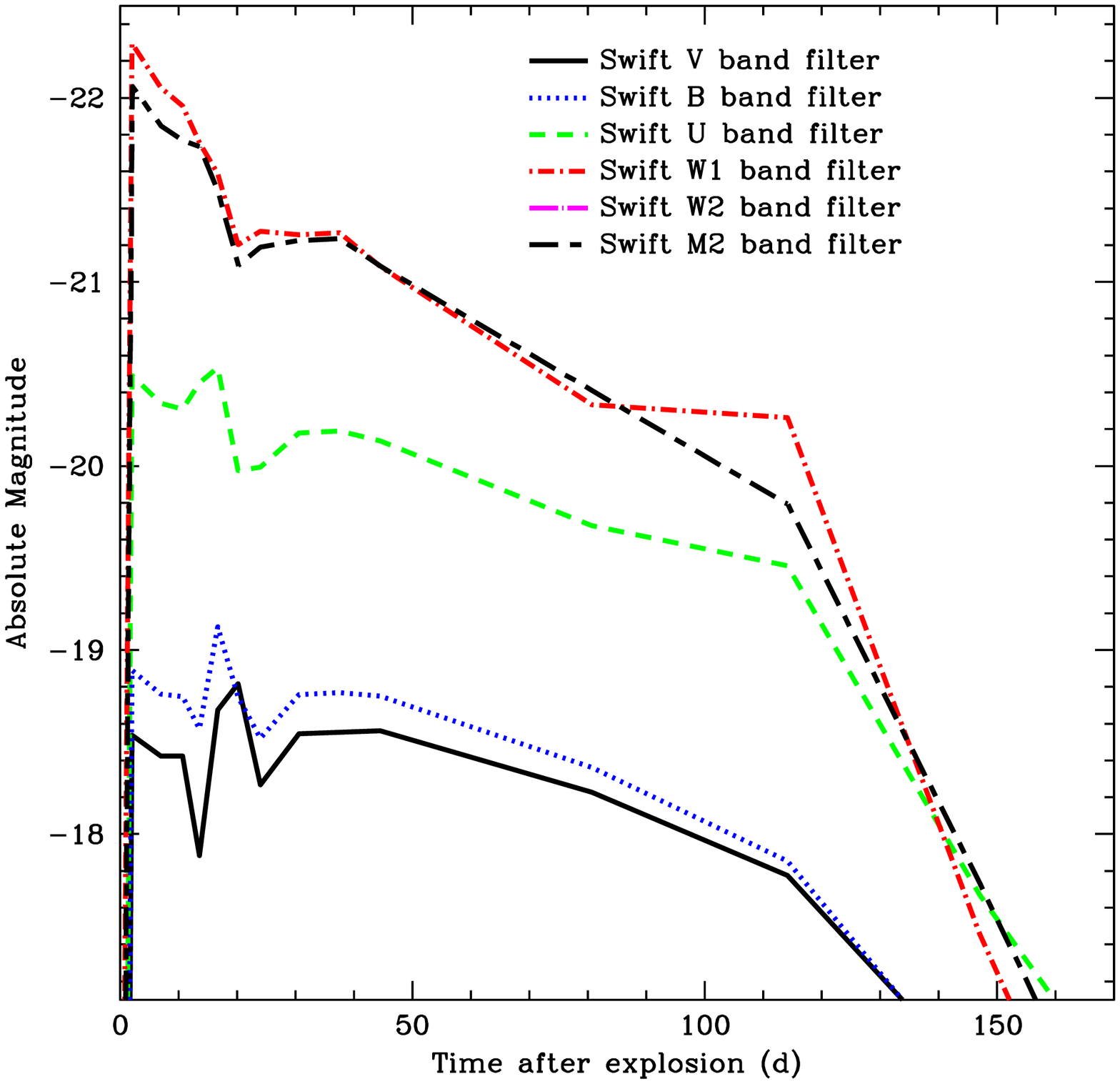}{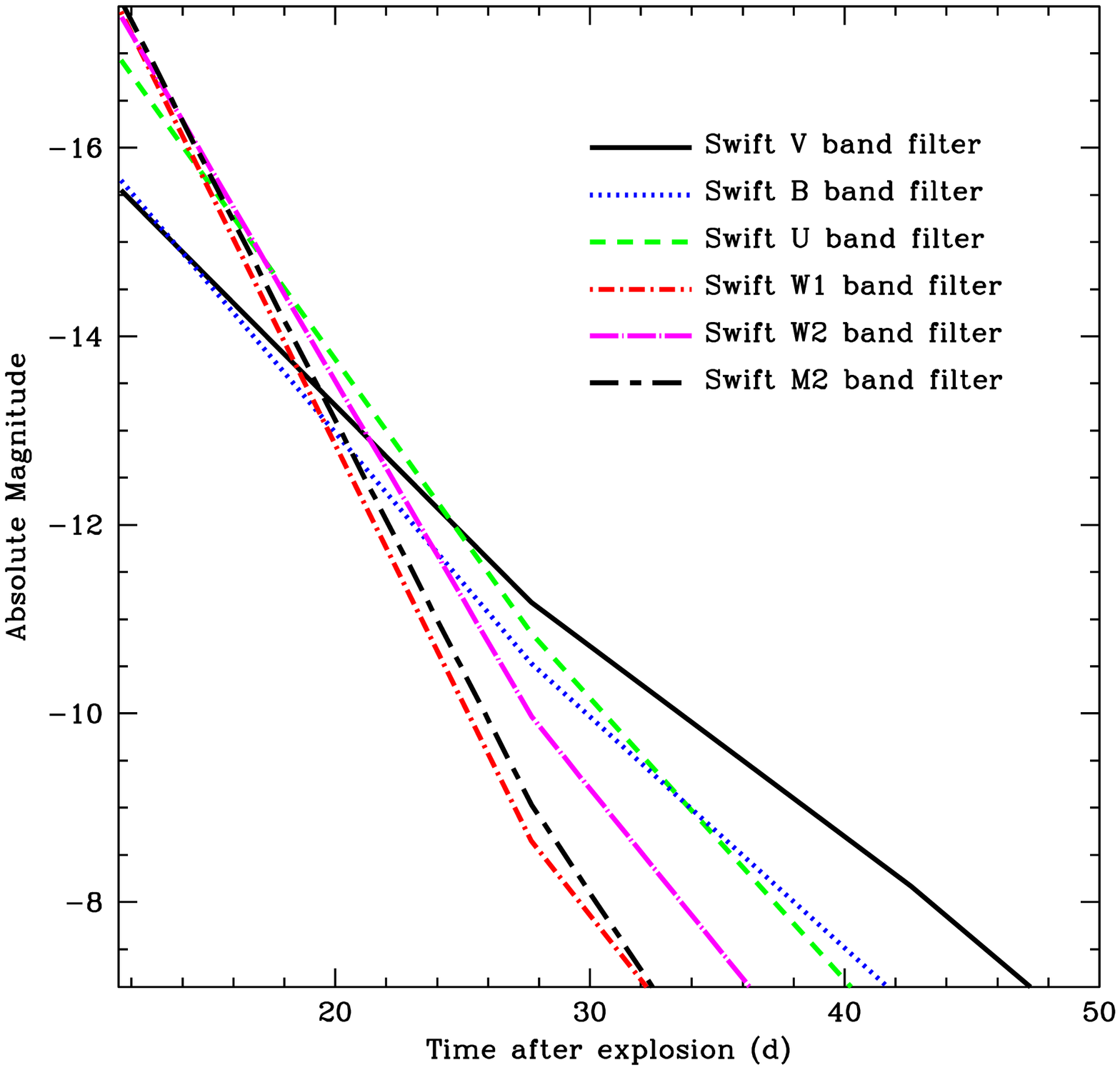}
\caption{The light curves of our .Ia supernovae (left: high density,
  right: low density).  Given the spectral peak in the X-ray
  (Fig.~\ref{fig:spec-.Ia}), it is not surprising that the denser
  (orbital plane) model is much brighter in the W1 and W2 bands than
  the V band.  Even so, in a dense medium, .Ia supernovae are nearly
  as bright as their Ia counterparts.  But in a diffuse medium (model
  based on binary interaction along the axis), the peak magnitudes are
  several magnitudes dimmer than type Ia supernovae.}
\label{fig:lc-.Ia}
\end{figure}

The emission from a .Ia supernova is strongly dependent on the density
profile.  First, there is very little radioactive ejecta, so it does
not contribute strongly to the light curve.  But with the fast ejecta
velocities and low masses, the gamma-rays from radioactive decay begin
to stream out early, also limiting how much the gamma-rays can
contribute to the light curve.  Recall that we deposit our energy
in-situ and hence we are overestimating the emission from this
low-density case.  The presence of fast moving ejecta means that
shocks are important for the light curve (Fig.~\ref{fig:lc-.Ia}).  For
our dense environment, the V and B bands peak between absolute
magnitudes of -18 to -19 (near to normal supernova brightnesses).  The
light curve will remain bright for nearly 100\,d.  The high shock
velocities and low densities lead to high temperatures and spectra
peaked in the UV and X-ray at early times (Fig.~\ref{fig:spec-.Ia}).
In the dense model, there is a decided drop in the UV band emission
after twenty days.  This occurs when the radiation leading the shock
emerges from ejecta and the radiation front leading the ejected shock 
cools.  From this point on, the photosphere of the explosion resides 
near the density peak in the ejecta.  Shock emergence is discussed 
in more detail by Frey et al. (in preparation).

But if the surrounding environment is diffuse, there is too little material 
to create strong emission, and the peak emission is limited to that 
of the initial explosion, peaking with absolute magnitudes around -16 
and dropping below -14 before 20\,d (below -12 by 30\,d).

The high temperatures and low densities limit the number of lines 
in the emitting regions and the regions just above these emitting 
regions.  Aside from absorption lines caused by the surrounding medium, 
we expect very few line features in their spectra.

\subsection{Neutrinos and Gravitational Waves from .Ia Supernovae}

.Ia supernovae will not be strong sources of neutrinos or GWs.

\subsection{Summary}

In this paper, we reviewed 3 separate explosions, providing the first 
radiation-hydrodynamics calculations of their emission (both spectra 
and light curves).  Each of these explosion scenarios will ultimately 
require detailed, individual studies as upcoming surveys begin to make 
first observations.  Here we have described many of the basic features 
one should expect in these explosions, focusing on the physics that 
alters the emission.

For supernovae, the emission is powered by a combination of the energy
from the decay of radioactive elements (primarily $^{56}$Ni and its
daughter products) and shock heating as the ejecta moves through its
surroundings.  For all of these ``failed supernovae'', the low $^{56}$Ni 
yield coupled to the high explosion velocities lead to peak emission 
dominated by the shock heating.  The energy from shock heating depends 
both on the velocity of the ejecta and the density structure of the 
surrounding medium.  For a given explosion, its surrounding medium 
(strength of the stellar wind, ejecta from a binary mass transfer 
phase, etc.) determines the peak luminosity and also shapes the 
spectra.  Unfortunately, neither the wind mass-loss nor the ejecta 
from binary mass transfer are well-known theoretically.  Observations 
of these transients will first and foremost help us in constraining 
the nature of these mass ejection mechanisms.

Depending upon the environment, the accretion-induced collapse
outburst could reach peak magnitudes that are nearly as bright as
normal supernovae, if only for a brief time (absolute magnitude in the
V band of -18.5, but dropping to below -17 by 30\,d).  If the total
mass ejecta is at the lower limit predicted by simulations, the peak
brightness will be several magnitudes dimmer than typical supernovae
(V band absolute magnitude of -16).  The rate of AICs could well be as
high as the type Ia supernova rate, but it could be several orders of
magnitude lower.  If typical peak magnitudes were at the higher range,
and the rate were truly close to the supernova rate, we should already
have observed some of these outbursts in existing samples.  The
progenitors of AICs are intimately linked to the progenitors of type
Ia supernovae, and understanding the AIC progenitor will teach us
about this supernova progenitor.  A nearby detection of an AIC will
also provide insight into neutron star formation.  Without the
convective engine and possibility of fallback, the neutrino signal
from AICs is a pristine measurement of stellar collapse and bounce.  We
can use the neutrino signal to study nuclear physics and the formation
of the neutron star.  In addition, because of the potential for rapid
spinning prior to collapse, AICs are the most promising candidates for
bar mode, and related, instabilities in the proto-neutron star, a
strong source of GW emission.  The observation, or lack thereof, of a
GW signal can be used both to understand the white-dwarf accretion
process and the nature of these proto-neutron star instabilities.

Type .Ia supernovae eject even less $^{56}$Ni (and less total mass)
than our AIC model, but the explosive energies are higher.  Shock
heating will dominate this explosion's emission, and hence the
emission from these explosions depends even more strongly on the
surrounding medium.  The peak V-band absolute magnitudes ranged from
-18.5 (holding above -18 for nearly 100\,d for dense surroundings) to
-16 (dropping below -12 in 25\,d).  Clearly, if the former were true,
these should have been observed in our current transient surveys and
it is likely that the answer lies somewhere in between these two
extremes.  Observations of type .Ia supernovae will place constraints
on binary mass transfer processes, ultimately improving our
understanding of this process.  This, in turn, will teach us 
about the progenitor scenarios for type Ia supernovae.

Fallback in supernovae, preferentially occuring in weaker explosions,
can drastically decrease the $^{56}$Ni yield.  We have studied an
extreme case where fallback ultimately causes the core to collapse to
form a black hole.  With its low $^{56}$Ni yield yield and low ejecta
velocities, it is the dimmest of all our models.  For the object
studied in this paper that produced a 4.5\,M$_\odot$ black hole, the
brightest explosion (with a Wolf-Rayet wind medium) peaks at V-band
absolute magnitudes of -15, dropping below -12 after 40\,d.  Such
systems would be very difficult to observe, but they are likely to be
the most common in the formation of stellar-massed black holes.  Their
neutrino signals would have delayed emission arising from a long
convective stage and fallback as material accretes onto the
proto-black hole.  These systems would also exhibit oscillations in
the general relativity metric and are prime sites to observe GWs from
black hole ringing.

\acknowledgements This work was carried out in part under the auspices
of the National Nuclear Security Administration of the U.S. Department
of Energy at Los Alamos National Laboratory and supported by Contract
No. DE-AC52-06NA25396.

\clearpage

\begin{deluxetable}{lcccc}
\tablewidth{0pt}
\tablecaption{Transient Models}
\tablehead{
  \colhead{Name}
& \colhead{Total Energy}
& \colhead{Nickel Yield}
& \colhead{Total Mass} 
& \colhead{Peak V}\\
  \colhead{}
& \colhead{$10^{51}$\,ergs}
& \colhead{M$_\odot$}
& \colhead{M$_\odot$}
& \colhead{magnitude}
}

\startdata

AIC & 2.0 (0.2\tablenotemark{a}) & 0.041 (0.0041) & 0.1925 (0.01925) &-18.5 (-16) \\ 
Type .Ia & 7.0 & 0.014 & 0.10 & -16 to -19 \\ 
Fallback & 1.7 & $10^{-13}$ & 3.0 & -13 to -15 \\

\enddata

\tablenotetext{a}{For our AICs, we have high and low ejecta models.}

\label{table:cond}
\end{deluxetable}

\clearpage

\clearpage

\clearpage

\clearpage

\clearpage

\clearpage

\clearpage

\clearpage

\clearpage

\clearpage

\clearpage

\clearpage

\end{document}